\documentclass[12pt,indent]{article}
\usepackage{a4,latexsym,amsmath,amssymb}
\usepackage[latin1]{inputenc}
\usepackage{float}

\usepackage{booktabs,caption}
\usepackage[flushleft]{threeparttable}
\usepackage{rotating}

\usepackage{authblk} 
\usepackage{natbib}
\usepackage{graphics}
\usepackage[english]{babel}
\usepackage{tabularx}
\usepackage{lscape}
\usepackage{multirow}
\usepackage{rotating}
\usepackage{dsfont}
\usepackage{caption}
\usepackage{subcaption}
\usepackage{bbm}
\usepackage[table]{xcolor}
\usepackage{booktabs}
\usepackage{calc}
\usepackage{ifthen}
\usepackage{url}
\usepackage{colortbl}      
\usepackage{tikz}

\usepackage{blindtext}
\usepackage{scrextend}
\usepackage{mathtools}
\usepackage{verbatim}
\usepackage{amsthm} 
\theoremstyle{definition}

\usetikzlibrary{shapes,snakes,matrix}

\newenvironment{tabularsmall}
{ \footnotesize \sffamily \tabular } {
\endtabular
\normalfont }

\newcommand{\E}{\operatorname{E}}      
\newcommand{\var}{\operatorname{var}}

\newcommand{\gammab}{\boldsymbol{\gamma}}

\newcommand{\xb}{\boldsymbol{x}}

\newcommand{\blanco}[1]{}

\def\d{\displaystyle}

\usepackage{natbib}

\usepackage{geometry}
\usepackage{setspace}
\usepackage{tikz}
\usepackage[]{graphicx}
\usepackage[]{color}
\makeatletter
\def\maxwidth{ %
  \ifdim\Gin@nat@width>\linewidth
    \linewidth
  \else
    \Gin@nat@width
  \fi
}
\makeatother

\definecolor{fgcolor}{rgb}{0.345, 0.345, 0.345}

\usepackage{framed}
\makeatletter
 {\par\unskip\endMakeFramed%
 \at@end@of@kframe}
\makeatother

\definecolor{shadecolor}{rgb}{.97, .97, .97}
\definecolor{messagecolor}{rgb}{0, 0, 0}
\definecolor{warningcolor}{rgb}{1, 0, 1}
\definecolor{errorcolor}{rgb}{1, 0, 0}

\usepackage{alltt}
\IfFileExists{upquote.sty}{\usepackage{upquote}}{}

\newtheorem{theorem}{Proposition}[section]

\begin{document}
\bibliographystyle{chicago}
\sloppy

\makeatletter
\renewcommand{\section}{\@startsection{section}{1}{\z@}%
        {-3.5ex \@plus -1ex \@minus -.2ex}%
        {1.5ex \@plus.2ex}%
        {\reset@font\large\sffamily}}
\renewcommand{\subsection}{\@startsection{subsection}{1}{\z@}%
        {-3.25ex \@plus -1ex \@minus -.2ex}%
        {1.1ex \@plus.2ex}%
        {\reset@font\normalsize\sffamily\flushleft}}
\renewcommand{\subsubsection}{\@startsection{subsubsection}{1}{\z@}%
        {-3.25ex \@plus -1ex \@minus -.2ex}%
        {1.1ex \@plus.2ex}%
        {\reset@font\normalsize\sffamily\flushleft}}
\makeatother



\newsavebox{\tempbox}
\newlength{\linelength}
\setlength{\linelength}{\linewidth-10mm} \makeatletter
\renewcommand{\@makecaption}[2]
{
  \renewcommand{\baselinestretch}{1.1} \normalsize\small
  \vspace{5mm}
  \sbox{\tempbox}{#1: #2}
  \ifthenelse{\lengthtest{\wd\tempbox>\linelength}}
  {\noindent\hspace*{4mm}\parbox{\linewidth-10mm}{\sc#1: \sl#2\par}}
  {\begin{center}\sc#1: \sl#2\par\end{center}}
}


\def\R{\mathchoice{ \hbox{${\rm I}\!{\rm R}$} }
                   { \hbox{${\rm I}\!{\rm R}$} }
                   { \hbox{$ \scriptstyle  {\rm I}\!{\rm R}$} }
                   { \hbox{$ \scriptscriptstyle  {\rm I}\!{\rm R}$} }  }

\def\N{\mathchoice{ \hbox{${\rm I}\!{\rm N}$} }
                   { \hbox{${\rm I}\!{\rm N}$} }
                   { \hbox{$ \scriptstyle  {\rm I}\!{\rm N}$} }
                   { \hbox{$ \scriptscriptstyle  {\rm I}\!{\rm N}$} }  }

\def\d{\displaystyle}\def\d{\displaystyle}

\title{Latent Trait Item Response  Models for Continuous Responses }
\author[1]{Gerhard Tutz} 
\author[2]{Pascal Jordan}
\affil[1]{Ludwig-Maximilians-Universit\"{a}t M\"{u}nchen} 
\affil[2]{University of Hamburg}   

\blanco{
\author{\large{Gerhard Tutz} \\
\small{LMU Munich, Department of Statistics} \\
\small{Akademiestra{\ss}e 1, 80799 M\"{u}nchen, Germany}\\
\small{Email: tutz@stat.uni-muenchen.de }
\vspace{0.8cm}\\
\large{Pascal Jordan} \\
\small{University of Hamburg,  Institute of Psychology} \\
\small{Von-Melle-Park 5, 20146 Hamburg, Germany}
\small{Email: pascal.jordan@uni-hamburg.de }
}
}

\vspace{3cm}

\maketitle

\begin{abstract}  
\noindent
A general framework of latent trait item response models for continuous responses is given. In contrast to classical test theory models, which traditionally distinguish between true scores and error scores, the responses are clearly linked to latent traits. It is shown that classical test theory models can be derived as special cases but the model class is much wider. It provides, in particular, appropriate modelling of responses that are restricted in some way, for example, if responses are  positive or are restricted to an interval. Restrictions of this sort  are easily incorporated in the modeling framework. Restriction to an interval is typically ignored in common models yielding inappropriate models, for example, when modeling Likert-type data. The model also extends common response time models, which can be treated as  special cases. Properties of the model 
class are derived and the role of the total score is investigated, which leads to a modified total score. Several applications illustrate the use of the model including an example, in which covariates that may modify the response  are taken into account.
\end{abstract}


\newpage
\section{Introduction}

The development of item response theory that clearly separates the observed response  from the underlying latent traits has been mainly driven by 
the consideration of binary items, in which it is distinguished between a correct response and an incorrect one as outlined, for example, in \citet{rasch1960studies}. Items with continuous response formats have been largely considered within the framework of classical test theory, which distinguishes between a true score, essentially the expected response, and an error score \citep{lord2008statistical}. Although there are some approaches to modeling continuous responses there seems no general framework available that considers responses as generated by latent traits and item characteristics.    

Continuous responses occur in particular in the form of time to complete a task and responses within a line segment (e.g., responses on a visual analogue scale). Especially time response modeling, which has already been considered by \citet{rasch1960studies} has drawn much attention \citep{van2006lognormal,roskam1997models,ferrando2007item,boeck2004framework}.  
Also Likert scales, which are quite common in practice,  are often modeled as continuous responses despite of an ongoing discussion as to whether that is the right way to analyse this type of data. An overview concerning problems and the pros and cons has been given by \citet{harpe2015analyze}. The controversy focuses on the problem if  Likert-type categories constitute interval-level measurement or have to be treated as ordered responses. If the  measurement level is only ordinal, then it is questionable if item sums should be used. \citet{harpe2015analyze} distinguish between the ``ordinalist'' and the ``intervalist'' view and conclude 
that ``individual rating items with numerical response formats at least five categories in length may generally be treated as continuous data''. Though one has not to agree, it seems worthwhile to investigate if there is much difference between continuous and discrete modeling given  proper latent trait models for both cases are available.  

\begin{figure}[h!]
\centering
\includegraphics[width=7cm]{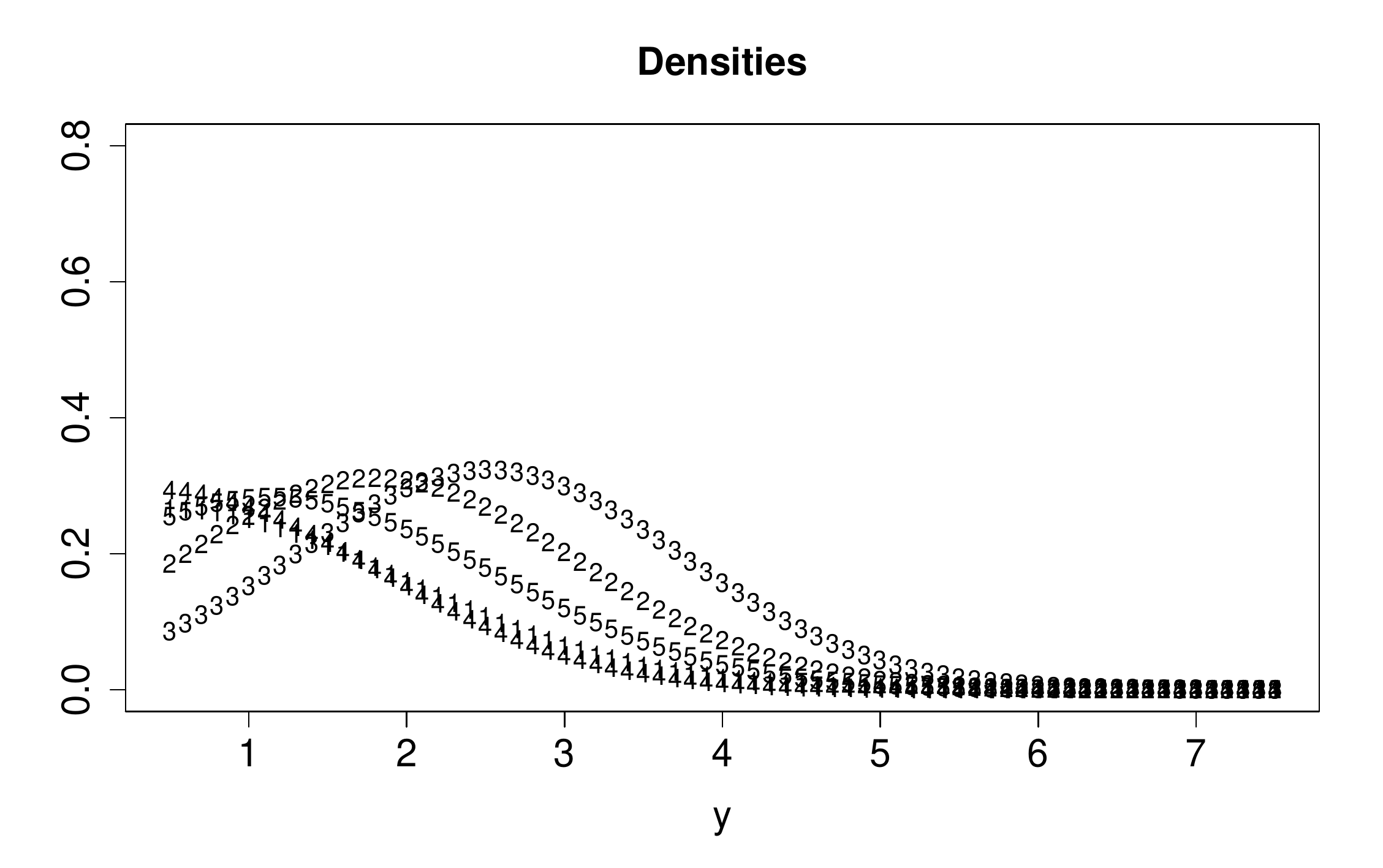}
\includegraphics[width=7cm]{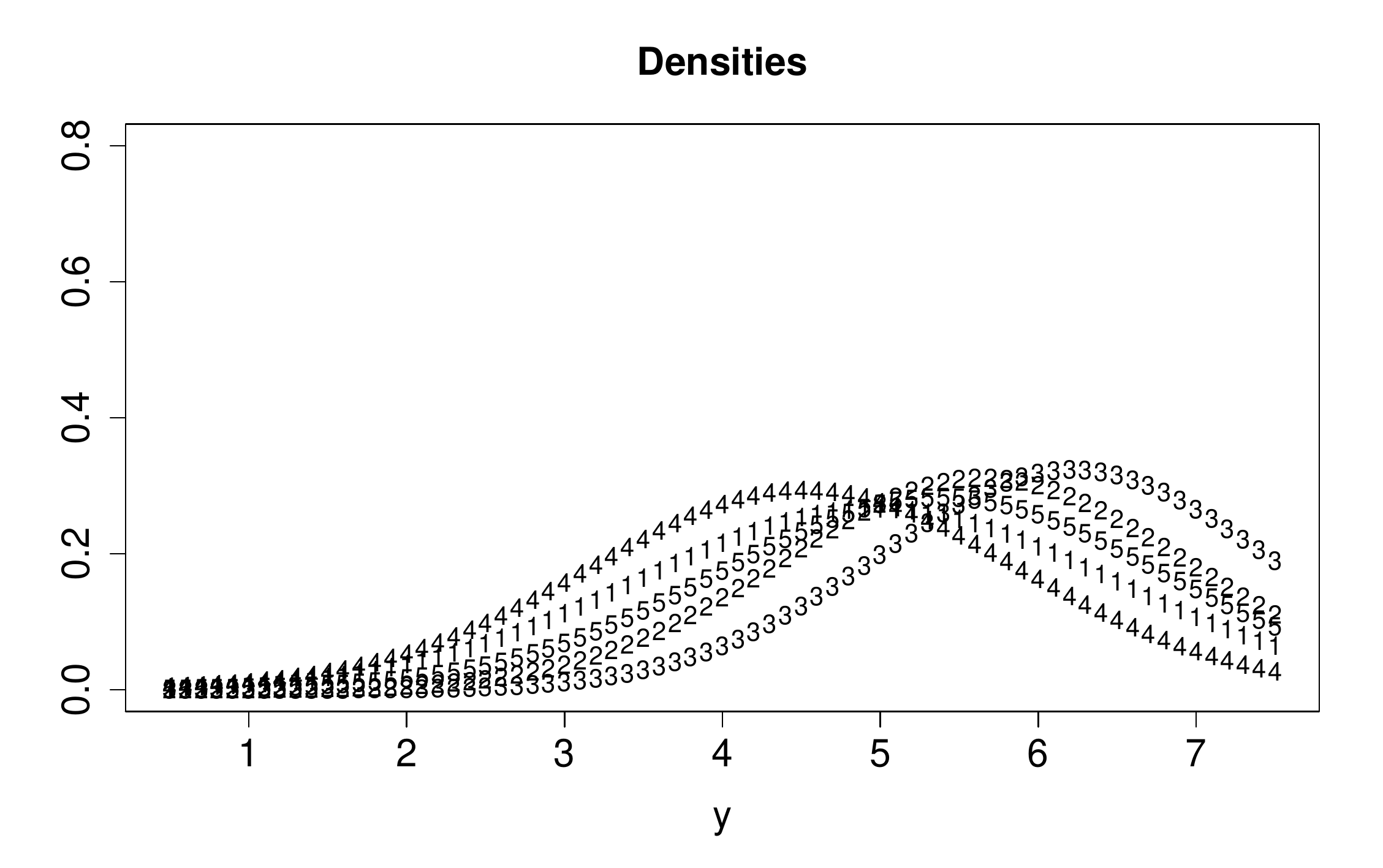}
\includegraphics[width=7cm]{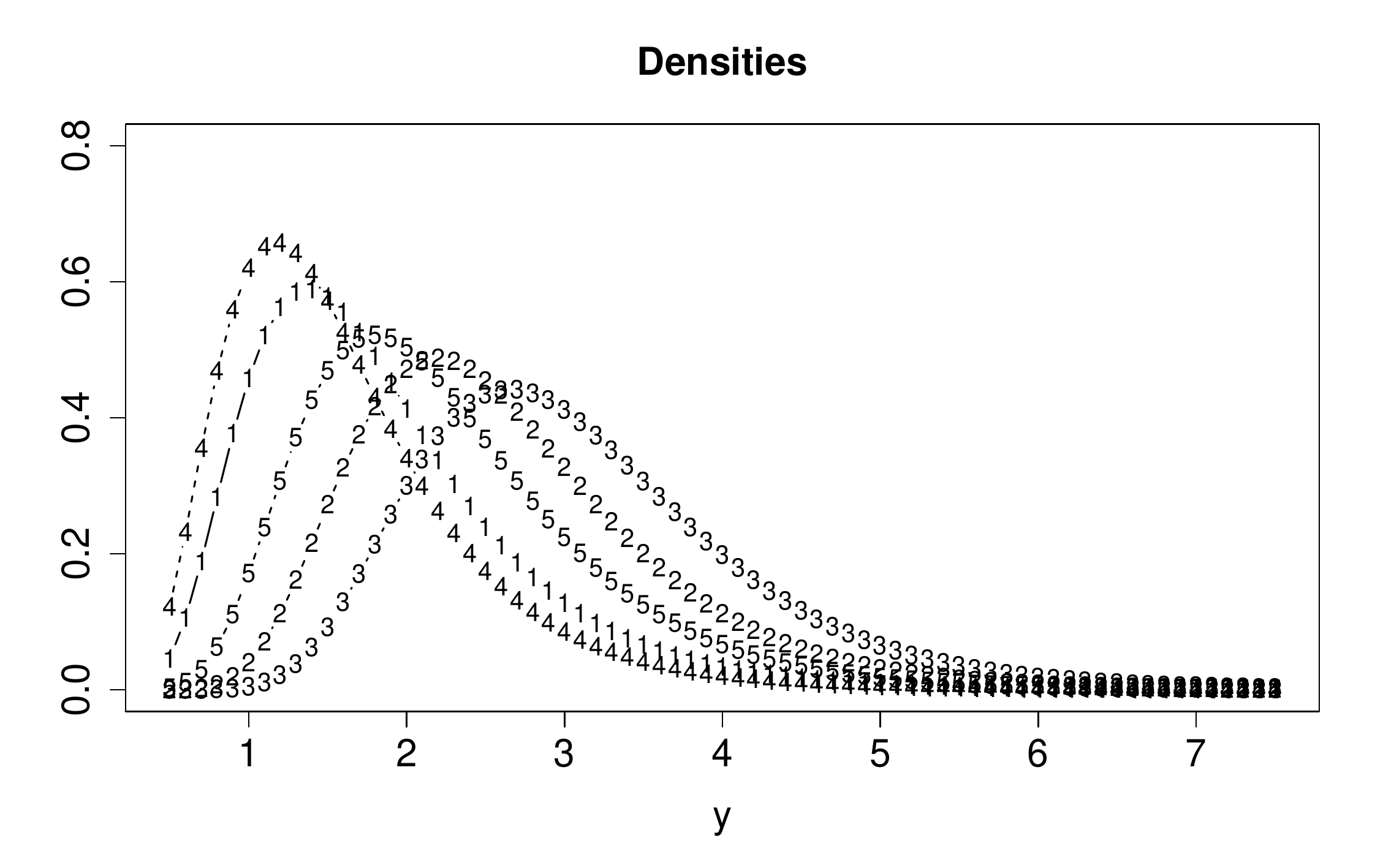}
\includegraphics[width=7cm]{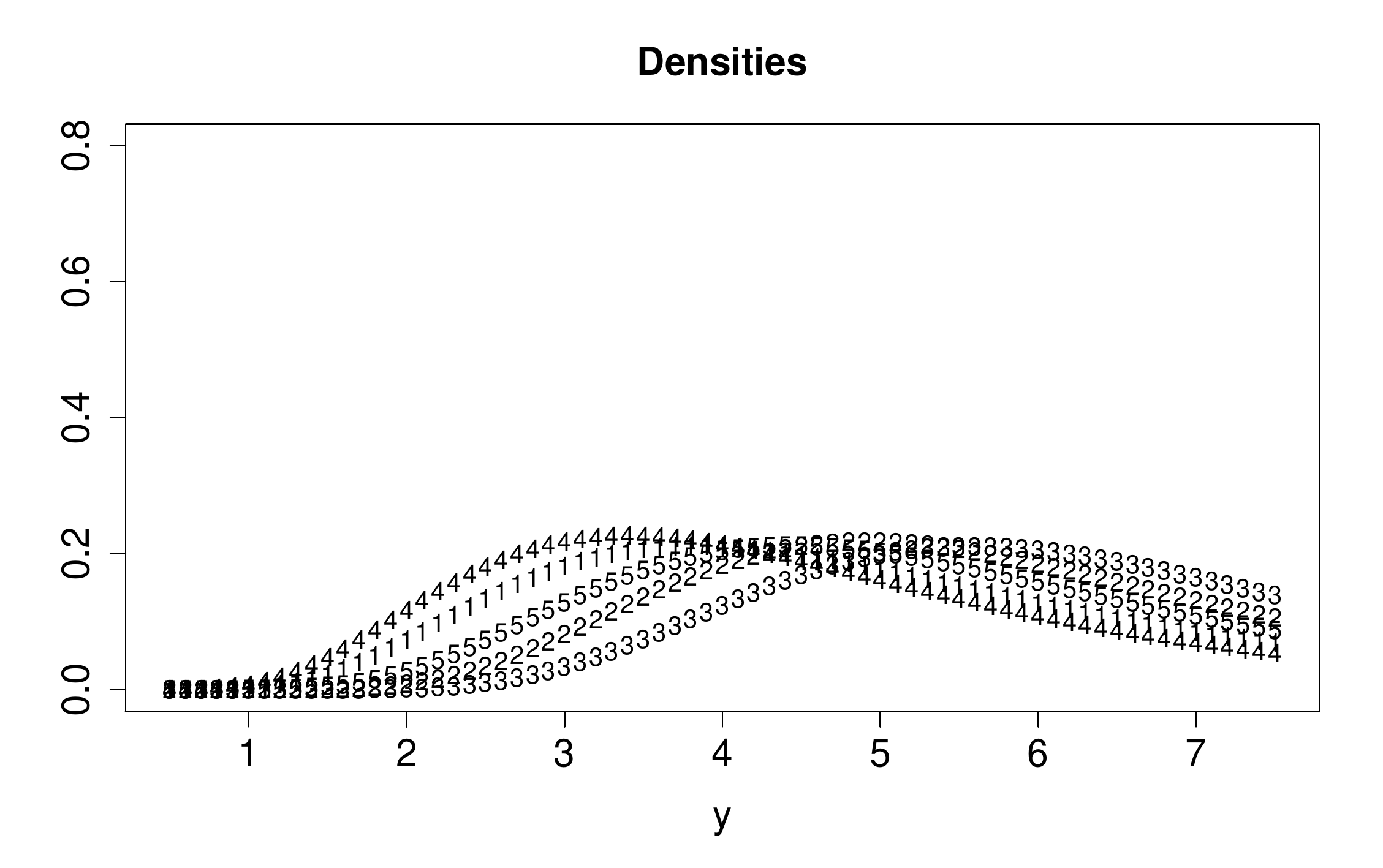}
\includegraphics[width=7cm]{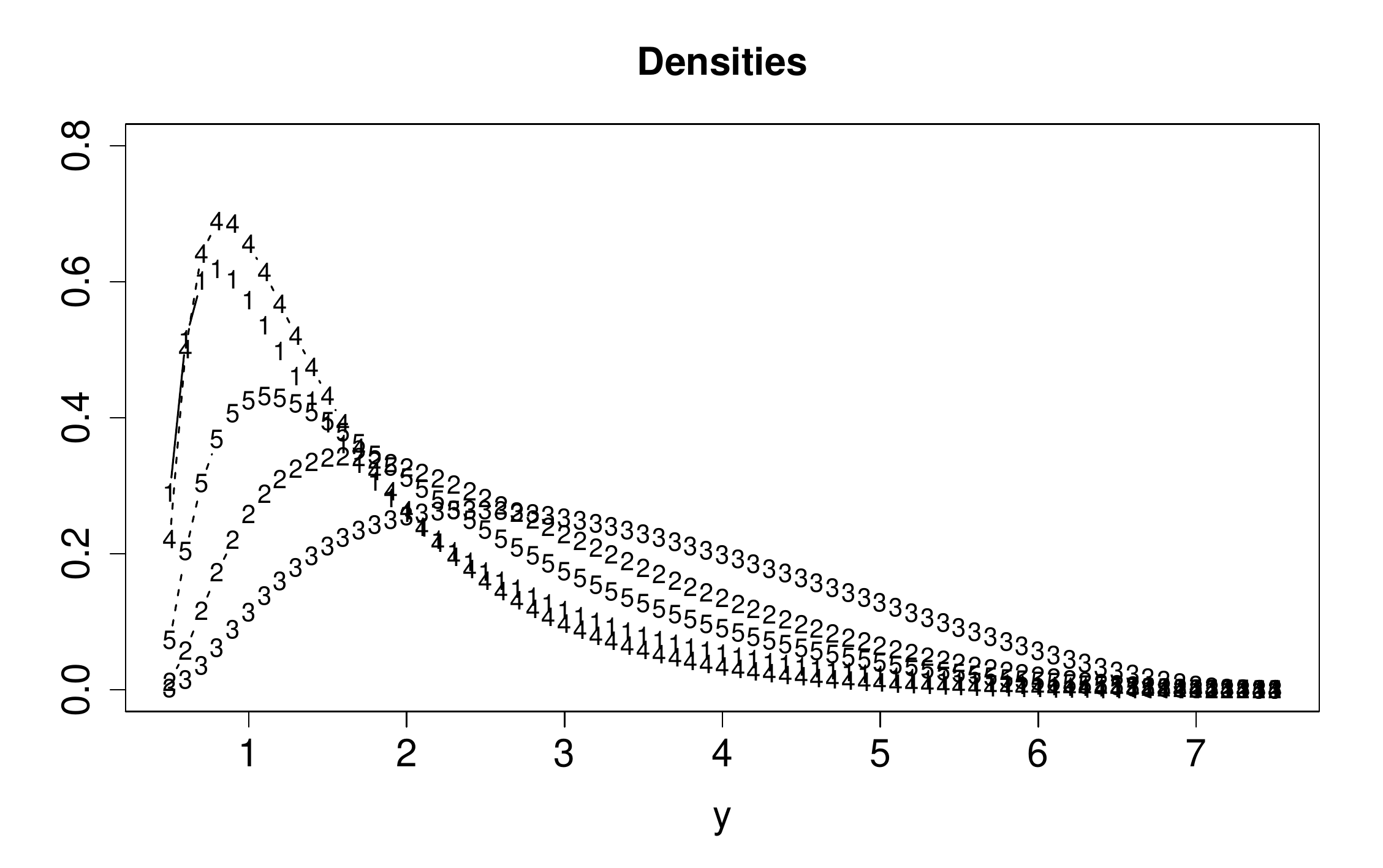}
\includegraphics[width=7cm]{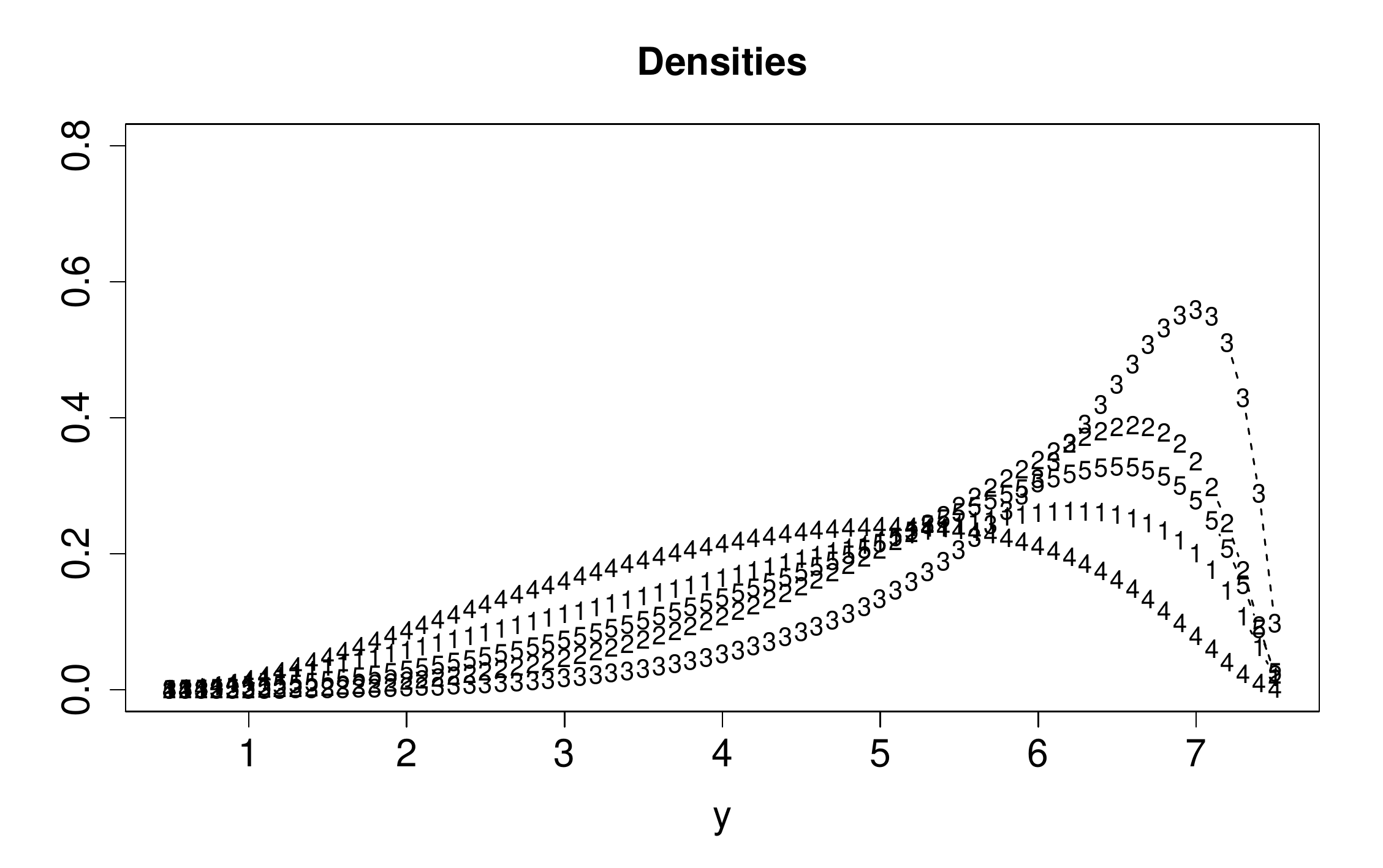}
\caption{Estimated densities for fear data, left: low person parameter, right: high person parameter, first row: normal distribution model corresponding to linear difficulty functions ($\theta_{\text{low}}=-2.5$, $\theta_{\text{high}}=0.5$)  second row: log-normal distribution corresponding to logarithmic difficulty functions ($\theta_{\text{low}}=-1.8$, $\theta_{\text{high}}=0.5$) third row: proper restrictions by using  logit difficulty functions $\theta_{\text{low}}=-1.8$, $\theta_{\text{high}}=0.5$).  In all cases the response function is fixed as the normal cdf.}
\label{fig:fearsill}
\end{figure}

When modeling Likert-type data, the main problem is that data are confined to an interval, which is typically ignored if one uses normal distribution models. For illustration Figure \ref{fig:fearsill} shows the resulting distributions for the fears data set to be considered in detail later (7-point Likert-type response) if one assumes the latent trait model considered here but with varying assumptions concerning the distribution of responses. The first row shows the fitted distributions if one assumes a normal distribution in a model that is equivalent to classical test theory. The left side shows the distributions for a low value of the person parameter, the right side shows the distributions for a high value of the person parameter. It is seen that the model does not yield proper distributions since the support of the distributions is much larger than the interval $(1,7)$. The same happens if one assumes a log-normal distribution (second row). Then the restriction to positive values is fulfilled but the upper boundary is ignored, which yields improper distributions.  The third row shows the fits for a latent trait model that explicitly accounts for the fact that responses are restricted to the interval $(1,7)$. Although we here considered Likert-type scales the problem is more general. The same inappropriate distributions occur in all cases in which  responses are restricted to an interval if one does not account for this restriction in a proper way.

The objective  of the present paper is to propagate  genuine latent trait models for continuous responses and investigate their properties.  A general framework for the IRT-based modeling of continuous data will be presented -- encompassing models for reaction times and visual analogue scales. Further, referring to the potential use in a Likert scale setting, we emphasize the need to properly account for the support of the data and show how this can be accomplished via the general modeling framework.  A special focus is on the link between the modeling class and classical test theory models.

The basic thresholds model approach has been outlined in \citet{TuItThres2021}. The present paper focuses on continuous responses and extends the approach in several ways. The link to various versions of the classical test theory model is investigated in detail, basic results are obtained by using quantile functions, which by itself show interesting properties of the model class, the role of total scores is examined and modified versions are proposed. Also the embedding of response time models, the comparison of models with differing response functions and the explicit inclusion of explanatory variables have not been considered before.

The paper is structured as follows: we first present the general modeling framework in Section 2. As a special case, we discuss linear models in Section 3, where we also outline relations of the proposed modeling class to classical test theory. The practically important cases of nonlinear models (encompassing e.g. reaction time data) will be treated in Section 4. In the latter section, we also expand  on how to account for restrictions regarding the support of the data. Also continuous approximations for Likert scale data will   be discussed in this section. Finally, we illustrate the  proposed modeling approaches via various datasets with differing response formats, including reaction times and Likert scales. In the latter case, the threshold model should be seen as an improvement over the default normal model which does not account for any restrictions in the support of the data.   

%

\section{Thresholds Models: Basic Concepts}\label{sec:basic}

Let $Y_{pi}, p=1,\dots,P, i=1,\dots,I, $ denote the responses of person $p$ on item $i$  having support $S_i$. As is common in IRT, we assume local stochastic independence (LSI) between the responses on different items given $\theta_p$. The general threshold model \citep{TuItThres2021} is given by   
\begin{equation}\label{eq:thr}
P(Y_{pi} > y|\theta_p,\alpha_i,\delta_{i}(.))=F(\alpha_i(\theta_p-\delta_{i}(y))),
\end{equation} 
where $F(.)$ is a strictly monotonically increasing, fixed distribution function, $\theta_p$ a person parameter, $\alpha_i$ a strictly positive discrimination parameter, and $\delta_{i}(.)$ a non-decreasing item-specific function, called \textit{item difficulty} function, which is defined on the support $S_i$. 
The function $F(.)$ is a \textit{response function}, which in combination with the difficulty function determines the distribution of the response. The form of the model reminds of binary response models as the normal-ogive or the two parameter logistic model, which are indeed   special cases if one considers  discrete responses $Y_{pi} \in \{0,1\}$, and defines the item difficulty parameter by $\delta_{i}=\delta_{i}(0)$ (and additionally setting $\delta_i(1)=\infty$).
However, in the present paper we confine ourselves to metric responses.

One of the important features of the model is that  $F(.)$ is a strictly increasing distribution function. Therefore, for fixed threshold $y$ the probability of a response larger than $y$ increases with increasing person parameter $\theta_p$, which makes the model a sensible latent trait model -- fulfilling the properties of a monotone homogeneity model \citep{sijtsma2002introduction} and thereby allowing nonparametric tests of the implied conditional association assumption \citep{holland1986conditional}.
The parameter  $\theta_p$ can be seen as  an ability or attitude parameter, which indicates the tendency of a person to obtain a high score. Higher values of $\theta_p$ are associated with a greater chance of scoring above some threshold $y$ for each item. For more details on the general  model see \citet{TuItThres2021}.

The concrete form of the thresholds model is determined by the choice of the difficulty functions $\{\delta_i(.)\}_{i=1, \dots, I}$ and the response function $F$. The model can be abbreviated by  TM($F$,$\{\delta_i(.)\}$).   If all the $\delta$- functions have the same structure,   instead of explicitly giving the functions we will use, for example, TM(normal,linear) if $F(.)$ is the (standard) normal distribution function, and the difficulty is linear, i.e. of the form $\delta_i(y)=\delta_{0i}+\delta_{i}y$.

\section{Linear Models}\label{sec:lin}

We will first consider models in which the mean of the response is a linear function of the latent ability. This can be obtained within the framework of  thresholds models by assuming that the  difficulty functions are linear. An important feature is that any strictly continuous response distribution  can be obtained by combining a linear difficulty function with a response function that is chosen according to the assumed response. 

\subsection{Linking Latent Traits and Responses}

Let $F(.)$ denote a fixed, typically standardized,  distribution function with support $\mathbb{R}$, for example the standardized normal distribution function, and $f(y)=\partial F(y)/ \partial y$ be the corresponding density. In addition, let the difficulty function be linear, $\delta_{i}(y)= \delta_{0i}+ \delta_i y$, $\delta_i >0$. Thus, we are considering threshold models of the type TM($F$,linear).

When investigating the distribution of responses it is helpful to define the distribution function $\bar F(y)=1-F(-y)$. If a random variable $Y$ has distribution function $F(.)$, the random variable $-Y$ has distribution function $\bar F(.)$. With $\bar f(y)=\partial \bar F(y)/ \partial y$ denoting the  density corresponding to $\bar F(.)$ 
one obtains for the distribution and the density of responses 
\begin{align}
&F_{pi}(y)=P(Y_{pi}\le y) = 1- F(\alpha_i(\theta_p-\delta_{0i}-\delta_{i}y))= \bar F(\alpha_i(\delta_{0i}+\delta_{i}y-\theta_p)) \nonumber\\
&f_{pi}(y)=\partial F_{pi}(y)/ \partial y= f(\alpha_i(\theta_p-\delta_{0i}-\delta_{i}y))\alpha_i\delta_{i}= \bar f(\alpha_i(\delta_{0i}+\delta_{i}y-\theta_p))\alpha_i\delta_{i}, \label{density}
\end{align}
That means the distribution function of the responses, $F_{pi}(y)$, is a shifted and scaled version of $\bar F(.)$, with the shifting and scaling depending on the person and the item.

The expectation and  variance of $Y_{pi}$ have the form (Proposition \ref{meanmoments} in the appendix) 
\begin{align}
&\mu_{pi}=\E(Y_{pi}) = \frac{\theta_p-\delta_{0i}-\mu_F/\alpha_i}{\delta_i}=  \gamma_i\theta_p - \gamma_{0i}, \label{condex}\\
&\sigma_{pi}^2=\var(Y_{pi}) =   \frac{\var_F}{\alpha_i^2 \delta_i^2}, \label{condvar}
\end{align}
where $\gamma_i= {1}/{\delta_i}$, $\gamma_{0i} =(\delta_{0i}+\mu_F/\alpha_i)/{\delta_i}$,  and  $\mu_F,\var_F$  are constants that are determined by the distribution function $F(.)$. More concrete,  $\mu_F=\int y f(y)dy$ is the expectation corresponding to distribution function $F(.)$ and  $\var_F =\sigma_F^2=\int (y-E_F)^2f(y)d y$ the corresponding variance. 
The main point is that the responses have a distribution function which is a shifted and scaled version of $\bar F(.)$, the means are linear functions of $\theta_p$, and the variances depend only on the items.
For symmetric distribution $F(.)$ one has simply $F(.)=\bar F(.)$, and $\mu_F=0$. Then responses follow the distribution function $F(.)$.

It is noteworthy  that one can choose any fixed function $F(.)$ (or $\bar F(.)$) and obtain   a  model in which responses follow the distribution function $\bar F(.)$ simply by using a linear difficulty function.  In particular,   one is not restricted to  normal distribution models, as is often done in applied research, but can try alternative distributions including non-symmetric ones. If, for example, one assumes for $\bar F(.)$ the Gumbel distribution, also known as maximum value distribution, $\bar F(y)= \exp(-\exp(-y))$, one obtains  for the responses a Gompertz distribution, $G(y)= 1-\exp(-\exp(y))$, if one assumes the Gompertz (or minimum value) distribution one obtains for the responses  the Gumbel distribution.

\begin{figure}[H]
\centering
\includegraphics[width=7cm]{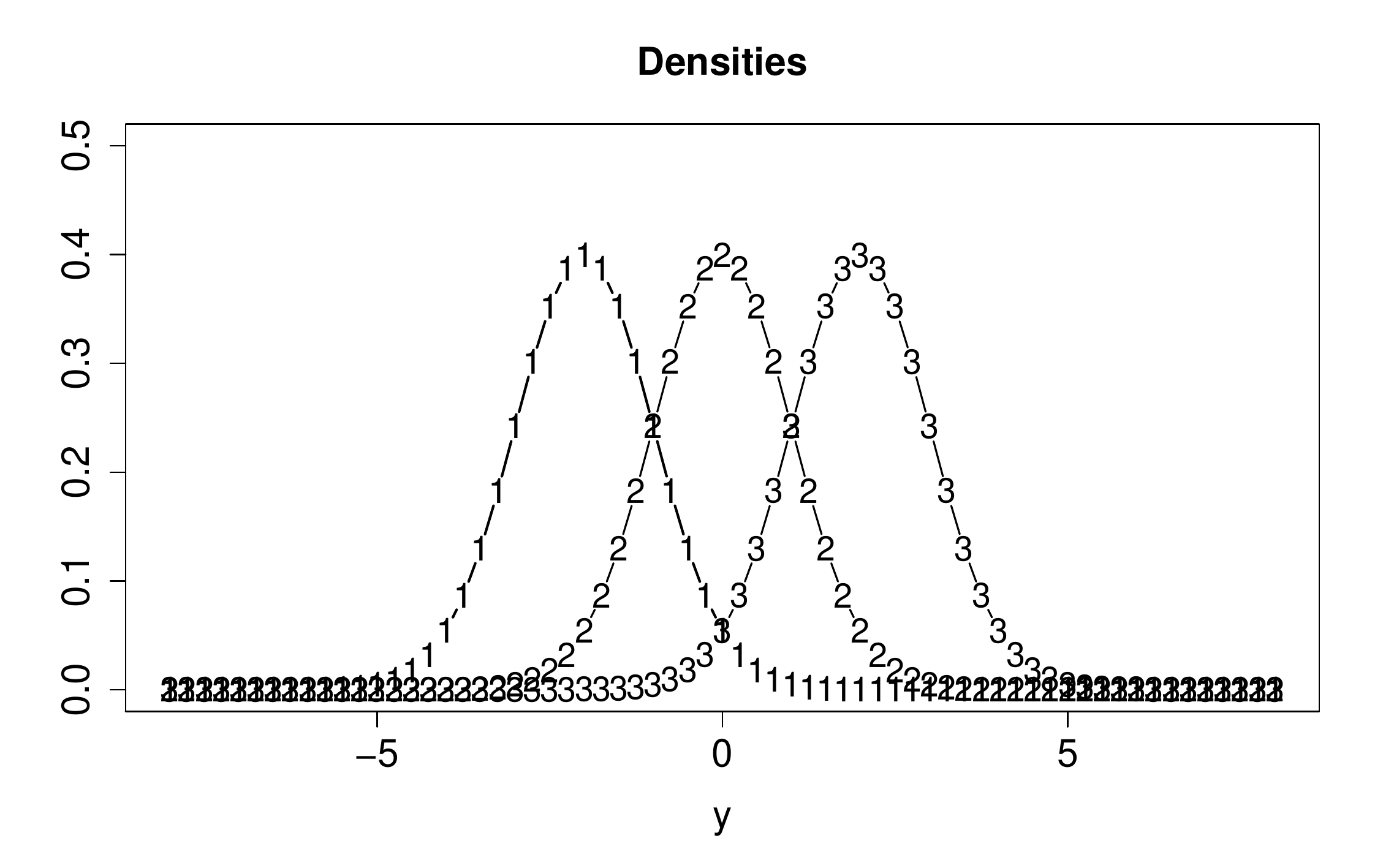}
\includegraphics[width=7cm]{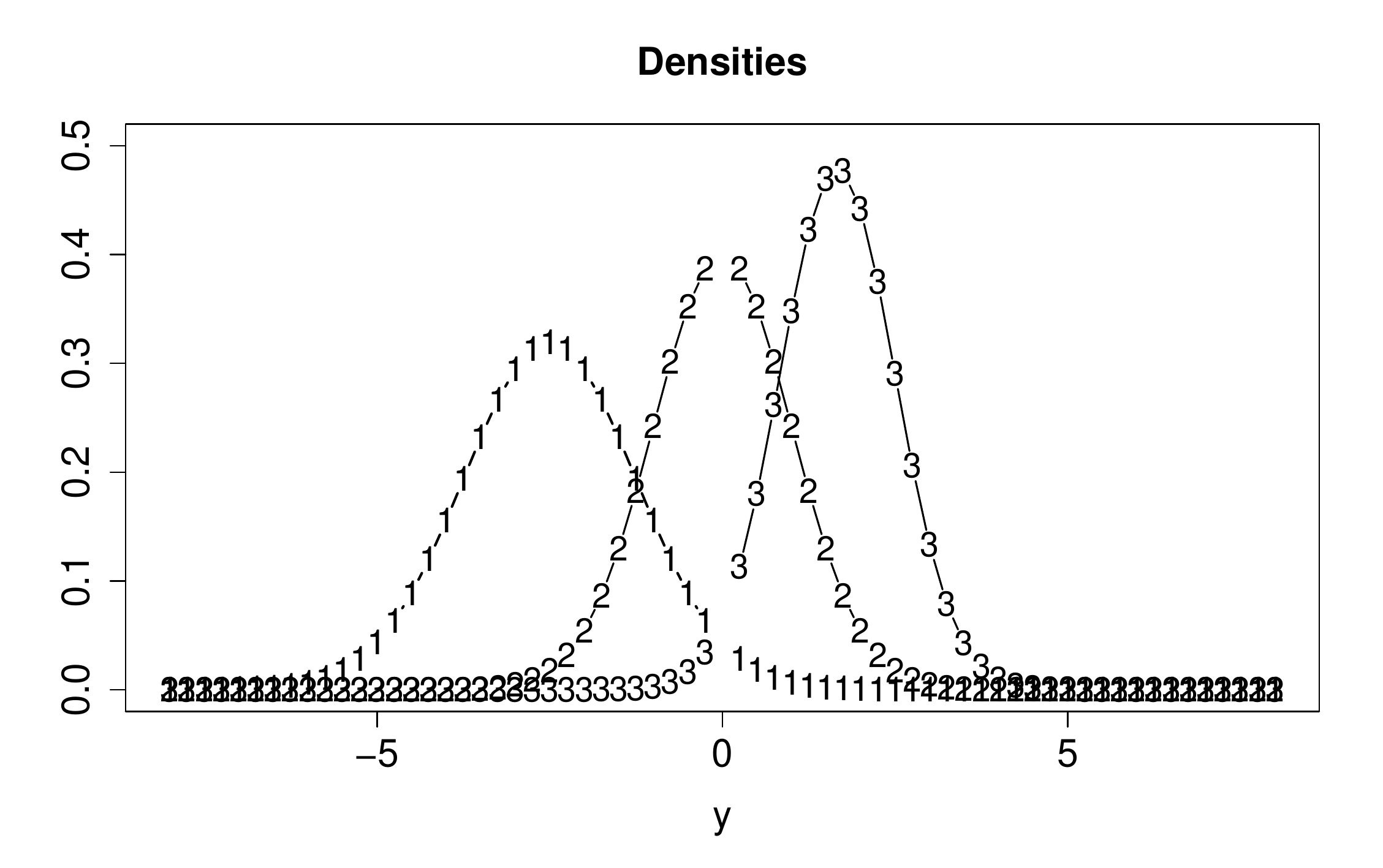}
\includegraphics[width=7cm]{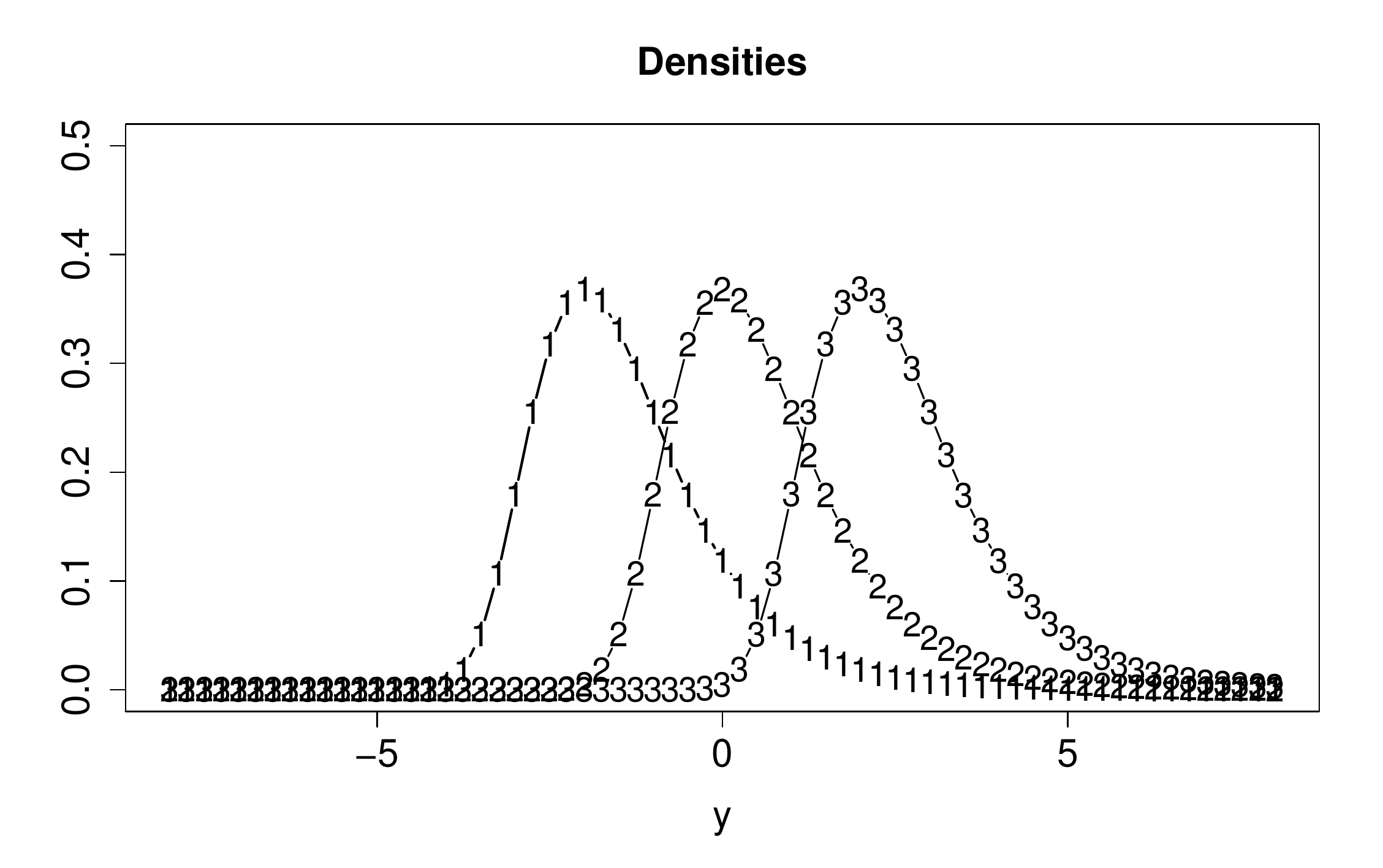}
\includegraphics[width=7cm]{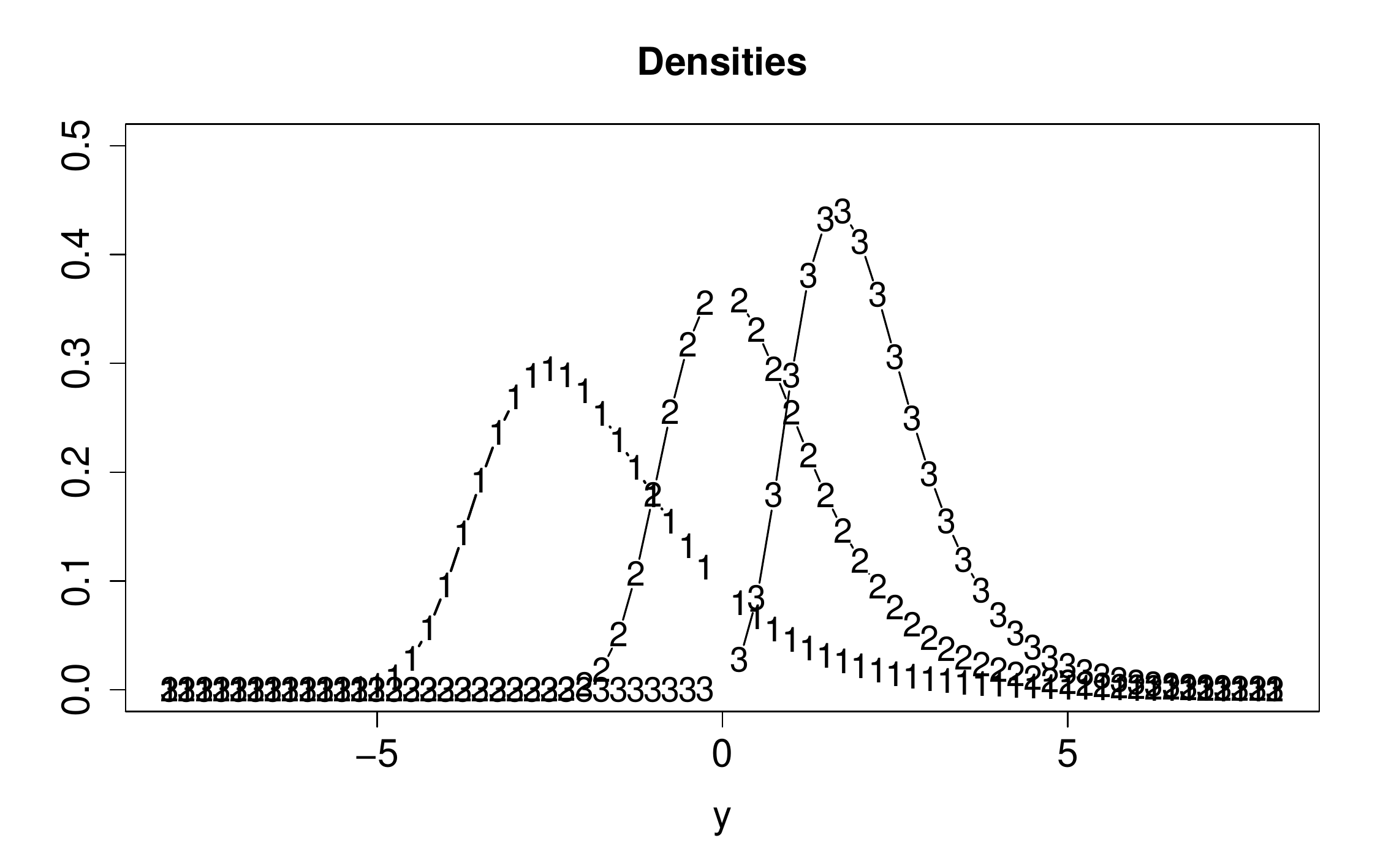}
\caption{Densities for three items with equal slopes and varying intercepts (first column) and for three items with varying slopes (second column). First row: normal response function, second row:  Gumbel for $\bar F(.)$.}
\label{fig:curves1}
\end{figure}

For illustration Figure \ref{fig:curves1} shows the distributions obtained for a person with $\theta=0$. The first row shows the densities if a normal distribution is assumed, the second row shows the densities if $F(.)$ is the standardized Gumbel distribution. The left picture show the results for three items  with  parameters $(2,1), (0,1), (-2,1)$ (for $(\delta_{0i},\delta_{i})$), in the right picture the slopes are varying with parameters $(2,0.8), (0,1), (-2,1.2)$. In all the pictures $\alpha_i=1$. If slopes are equal  (left picture) the intercepts determine the difficulty of the item, the highest responses are to be expected for item 3, which has the smallest intercept ($\delta_{0i}$). If slopes vary across items the mean as well as the variance change. Item 3 has the  smallest variance since it has the highest slope (right picture). The second row shows the corresponding densities if a Gompertz distribution is assumed for $F(.)$, consequently the responses follow the (skewed) Gumbel distribution. 

The derived results hold for all strictly continuous distribution functions $F(.)$, not only for functions with support $\mathbb{R}$.  Note however, that for distribution functions with a smaller support, the support of the item scores  typically differs for different latent abilities. If a response function with only positive support is chosen, that is, if $F(x)=0$ holds for $x<0$, then for fixed $y$, one may always find a large enough negative $\theta$, such that $(\theta-\delta(y))<0$ and hence $P(Y_{pi}>y)=0$ holds for that value of $y$. Whether this poses a practical problem depends on the probability distribution of the latent variable. Such a dependency of the support on $\theta$ is avoided if one restricts to response functions which are positive throughout $\mathbb{R}$ (as in the case of a normal or Gumbel distribution).

\subsection{Thresholds Models and Classical Test Theory}

In classical test theory (CTT),  the response usually is decomposed  into the ``true score'' and the ``error score'', typically at the population level. A similar decomposition holds for threshold models on the population level and  the person level. At the person level one has 
\begin{equation}\label{equ:decomp}
Y_{pi}= \tau_{pi}+ \varepsilon_{pi},
\end{equation}
where the noise variable  $\varepsilon_{pi}$ has expectation $\E(\varepsilon_{pi})=0$ and variance $\var(\varepsilon_{pi})={\var_F}/({\alpha_i^2 \delta_i^2})$ (the full derivation is given in Propostion \ref{ctt_theorem} in the appendix).
Moreover, the true-score $\tau_{pi}$ equals the expected value of $Y_{pi}$ and it depends on $p$ only via $\theta_p$, i.e. two test takers with the same latent ability possess the same true score.  

One can then define -- following \citet{novick1966axioms} and \citet{holland2003classical} -- the true score random variable $T_i$ as the true score of a randomly selected test taker from the population and the error variable $\varepsilon_i$ as the correspondingly sampled noise variable, when testing the selected test taker. That is, on the random sampling level, wherein $Y_i$ denotes the response of a randomly selected test taker on the $i$-th item, the following equation holds
\begin{equation}\label{equ:ktt}
Y_{i}= T_{i}+ \varepsilon_{i}.
\end{equation}
Herein, $T_i=T_i(\theta):=\mathbb{E}(Y_i|\theta)$ and $\varepsilon_{i}=\varepsilon_i(\theta):= Y_i - T_i(\theta)$ are functions of the random variable $\theta$. All central axioms of the CTT model \citep{novick1966axioms} are implied by the properties of a general (not necessarily linear) TM model -- as shown in Proposition \ref{ctt_theorem} in the appendix. Basically, these axioms boil down to the existence of  an additive decomposition (\ref{equ:ktt}) for each item, with the additional properties that $i)$ errors on different items are uncorrelated; $ii)$ errors and true scores are uncorrelated and $iii)$ the expectation of the error given the true score is zero.

This representation has several consequences. 
Given that a threshold model holds a CTT model holds for randomly selected test takers. Thus, all CTT based quantities, like reliability coefficient, may be defined appropriately and all of the derived results for true score prediction may be applied to the TM setting (for details we refer to \citet{holland2003classical}). Hence,  a plethora of already established results become applicable.  Another important aspect is that the threshold model can be seen as a latent trait model underlying the CTT model. In the   CTT model, expectations of item responses are simply considered as representing the  true scores,  but latent traits as the driving force behind an individual's responses  on items are not clearly identified.

Now, assuming the special case of a linear TM model, one may derive specific submodels of classical test theory. To this end, it is helpful to recall the following distinction \citep{raykov1997estimation}: Measurements are called \emph{congeneric} if all true scores may be expressed as affine functions of a single true score, i.e.,  
$T_i = a_i T + b_i$ holds for some fixed values $a_i,b_i$. This equals the notion of unidimensionality from a CTT point of view (providing also the decomposition  of the covariance matrix according to a one-dimensional factor analysis model, albeit lacking the independence assumptions). This property  is always satisfied for a \emph{linear} TM (see equation \ref{eq:linCT} below), but not necessarily for a general TM. 
For the linear model one obtains from  (\ref{condex})
\begin{align}\label{eq:linCT}
T_i(\theta)=\mathbb{E}(Y_i|\theta) = \frac{\theta -\delta_{0i}-\mu_F/\alpha_i}{\delta_i}
\end{align} 
and therefore
\[
\theta = \delta_i T_i(\theta) +\delta_{0i}+\mu_F/\alpha_i.
\]
Thus, measurements are congeneric. As will be shown in the following, in non-linear TMs $\theta$ is  a non-linear function, and measurements are not congeneric.
It should be noted that the two notions of unidimensionality (IRT vs. CTT) do not coincide. A TM model may be classified as a unidimensional IRT model according to common definitions \citep{holland1986conditional}, but if it is not a linear TM model, then the corresponding CTT model is not necessarily unidimensional (in the sense used in CTT modeling).

The latter is caused solely by the fact, that the CTT based definition of unidimensionality requires linear relationships on the level of the true scores, whereas a general TM model provides a nonlinear relation, as the following argument shows.   The dependency of the $i$-th true score on $\theta$ is represented by $T_i(\theta)=\mathbb{E}(Y_i|\theta)$. One may now fix some item $i$ (without loss of generality $i=1$) and express the latent ability as a function of the true score $T$ on that item via: $\theta(T)= T^{-1}_1(T)$, where $T$ denotes the true score on the first item (note that $T_1$ denotes the function, whereas $T$ is used to denote the true score variable). Substituting this expression for $\theta$, the true score on any other item can be expressed as a function of the true score of the reference item: $T_j(T):=T_j(\theta(T))$. In general, these functions differ from linearity and hence one does not arrive at a congeneric CTT model. 

One may further subdivide the congeneric model -- assuming in the following a linear TM. Measurements are called \emph{essentially tau-equivalent} if $a_i=1$ holds for all $i$ in the congeneric relationships $T_i=a_i T + b_i$ (or upon redefinition of $T$ we may just demand $a_i=c$ for some constant $c$). 
As is seen from (\ref{eq:linCT})
the model is  congeneric model with $a_i= {1}/{\delta_i}$ and $b_i =-\ ({\mu_F\alpha^{-1}_i+\delta_{i,0}})/{\delta_i}$.
This congeneric model is essentially tau-equivalent if $\delta_i=1$ holds for all $i$. Two important consequences of tau-equivalency shall be pointed out: 
Firstly, conditionally on $\theta$, the expected value of the unweighted mean of the item scores equals the latent variable plus a bias value (determined by the intercept term). As the bias term is independent of $\theta$, one may deduce that the difference of means of two test takers provides an unbiased estimate of their true difference in the ability -- thus justifying the usage of simple sum scores (although there are statistically more efficient estimators). Secondly, commonly applied coefficients -- such as Cronbach's alpha -- become reasonable estimators of the test reliability when tau-equivalency can be assumed \citep{jackson1977lower}.

 
The even stricter requirement of \emph{essentially parallel} measurements demands in addition to tau-equivalency the equality of the error variances.
From equation (\ref{condvar}), one obtains
\begin{align}\label{eq:condvar}
\text{Var}(\varepsilon_i)=&\text{Var}(\mathbb{E}(\varepsilon_i|\theta))+\mathbb{E}(\text{Var}(\varepsilon_i|\theta)) 
=\mathbb{E}(\text{Var}(Y_i|\theta)) = \frac{var_F}{\alpha^2_i \delta_i^2} 
\end{align}
Accordingly, if $\delta_i=1$ (which is necessary and sufficient for essentially tau-equivalency) holds all the error variances will be equal only if the discrimination parameters are also equal. \\
Two consequences of parallelity are worth mentioning. Firstly, the best linear predictor of the true score weights all item scores equally, providing further justification for the usage of simple sum scores. Secondly, estimation of test reliability via the split-half approach is justified.


%

Taken together, these results show that: 

$i)$ on the second order level (i.e. using only conditional expectations and variances) the general threshold model yields CTT models, 

$ii)$ the unidimensional CTT model can be motivated by an underlying linear thresholds models.  

\subsection{Quantile Function and Further Properties}
It has already been highlighted that the mean is a linear function of the latent ability and that the variance does not depend on $\theta$. But one may go further and examine the dependency of quantiles and quantile-based measures of spread on the latent abiliy. 

The quantile function for the response $Y_{pi}$ and values $0<q<1$ is given by
\[
Q_{Y_{pi}}(q) = \inf \{y|1-F(\alpha_i(\theta_p-\delta_{i}(y))) \ge q \}= \inf \{y|\delta_{i}(y) \geq \theta_p - F^{-1}(1-q)/\alpha_i \}.
\]
For strictly increasing difficulty functions $\delta(\cdot)$ mapping onto $\mathbb{R}$, which are assumed in the following, it has the simpler form 
\begin{align}\label{eq:quant}
Q_{Y_{pi}}(q) = \delta_{i}^{-1}(\theta_p - F^{-1}(1-q)/\alpha_i).
\end{align}
We now examine the case of linear difficulty functions more closely.
Since difficulty functions are linear one obtains $\delta_i^{-1}(x)= ({x-\delta_{0i}})/{\delta_i}$ and the quantile function reduces to
\[
Q_{Y_{pi}}(q) = \frac{\theta_p - F^{-1}(1-q)/\alpha_i -\delta_{0i}}{\delta_{i}}.
\]
Therefore, each quantile $Y_{pi}$ is a linear function of $\theta_p$. For any $q \in (0,1)$ the $q$-quantile increases linearly with $\theta_p$. A further consequence of the form of the quantile function is that common measures of spread that are based on quantiles, for example  the interquartile range $Q_{Y_{pi}}(.75)-Q_{Y_{pi}}(.25)$, do not depend on $\theta_p$.

The quantile function may be used to derive formulas for the moments of $Y_{pi}$. For the central moments one obtains 
\[
\E(Y_{pi}-\mu_{pi})^k=\int_0^1 \left(\frac{\mu_F-F^{-1}(1-q)}{\alpha_i \delta_{i}}\right)^k dq,
\]
see Proposition \ref{meanmoments}. A consequence is  that also central moments do not depend on $\theta_p$. It underlines that the person parameter just shifts the distribution of responses but does not affect its form, which is determined by the item parameters only. 

A simpler form of the density of responses may be obtained by rewriting (\ref{density}) in centered form as:
\begin{align}
f_{pi}(y) =& \bar f(\alpha_i(\delta_{0i}+\delta_i y -\theta_p))\alpha_i\delta_i = 
\bar f\left(\alpha_i\delta_i\left(y -\frac{\theta_p-\delta_{0i}}{\delta_i}\right)\right)\alpha_i\delta_i. \label{centered}
\end{align}
From the change of variable formula, one may deduce from (\ref{centered}) the following:\\
Let $X$ denote a random variable with distribution function $\bar F$. Then the random variable $Y:= aX+b$ with $a:= \frac{1}{\alpha_i \delta_i}$ and 
$b:= \frac{\theta_p-\delta_{0i}}{\delta_i}$ is distributed as $F_{pi}$.

A common measure for the performance of persons that is typically used is the total score $Y_{p+}=\sum_i Y_{pi}$. In linear models the
expectation and variance are given by
\begin{align*} 
\E(Y_{p+}) = \theta_p\gamma_{+}-\gamma_{0+},\quad
\var(Y_{p+}) =   \sum_i \frac{c}{\alpha_i^2 \delta_i^2}
\end{align*}
where $\gamma_{+}=\sum_i\gamma_i$, $\gamma_{0+}=\sum_i \gamma_{0i}$. Thus, for linear thresholds models the expected total score is essentially the latent score, and the variance does not depend on $\theta_p$.

\section{Non-linear Models}

In traditional item response models like the Rasch model or the normal-ogive model the mean response is a non-linear function of the  person's latent trait. This is sensible because the means in binary responses are restricted to the interval [0,1] and linear functions tend to take values outside this interval. In general, non-linear functions are always to be expected if the response is restricted in some way. This holds also if responses are continuous but restricted, for example to take positive values only, which is the case in many applications.

Within the framework of  thresholds models restrictions on the support of responses are obtained in a natural way by specifying appropriate non-linear difficulty functions. This leads to models in which the mean and other characteristics of the responses are non-linear functions of the latent trait. In the following we consider difficulty functions of the form $\delta_{i}(y)= \delta_{0i}+ \delta_i g(y)$, where $g(.)$ is a strictly increasing fixed function.

\subsection{Responses in the Positive Domain}

Let the response function $F(.)$ be chosen as fixed. The threshold model automatically restricts the responses to positive values, if for the difficulty function $\lim_{y\rightarrow 0}\delta_i(y)=-\infty$ holds. One candidate is the logarithmic function $g(y)=\log(y)$ yielding $\delta_{i}(y)= \delta_{0i}+ \delta_i \log(y)$.

\begin{figure}[H]
\centering
\includegraphics[width=7cm]{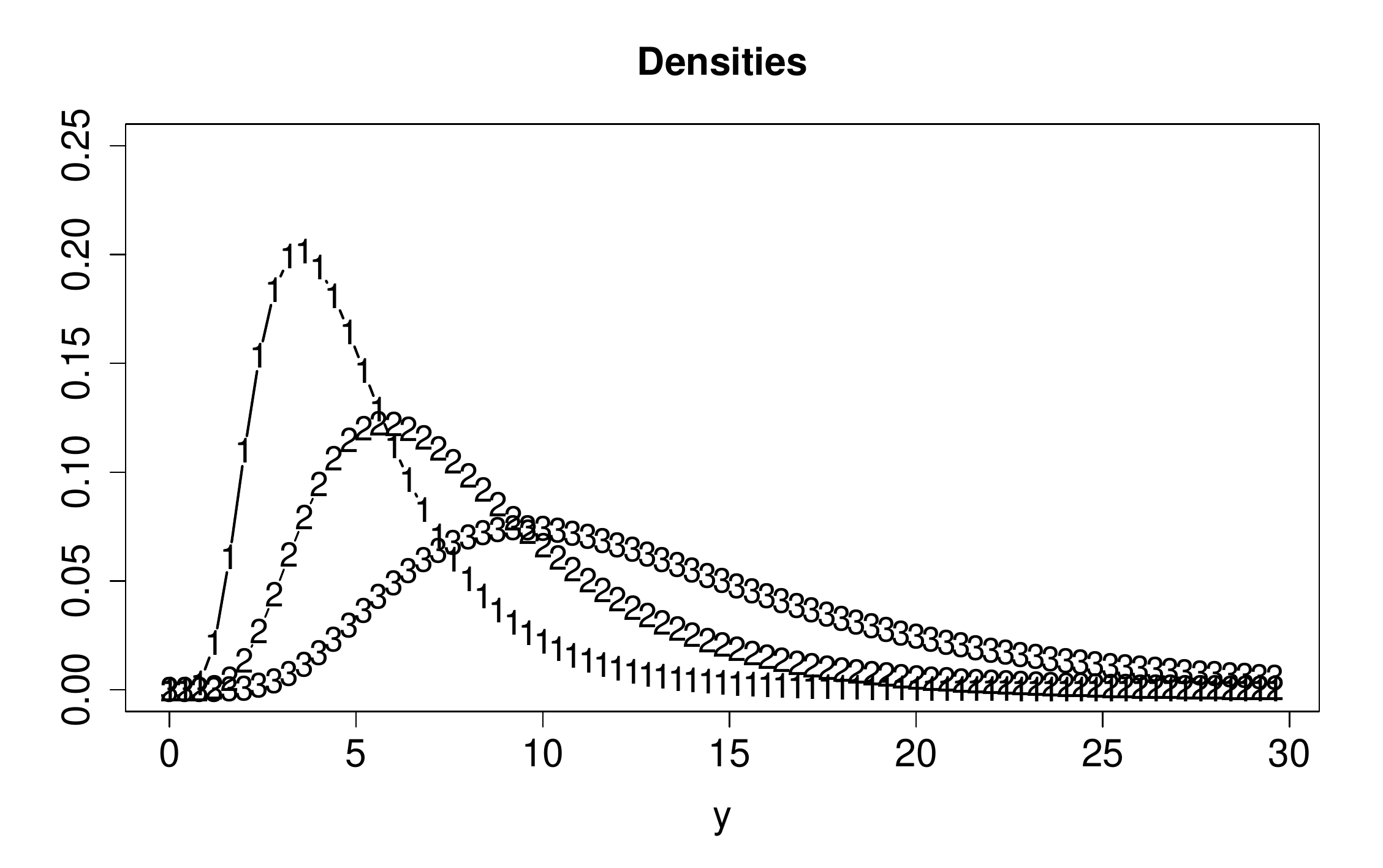}
\includegraphics[width=7cm]{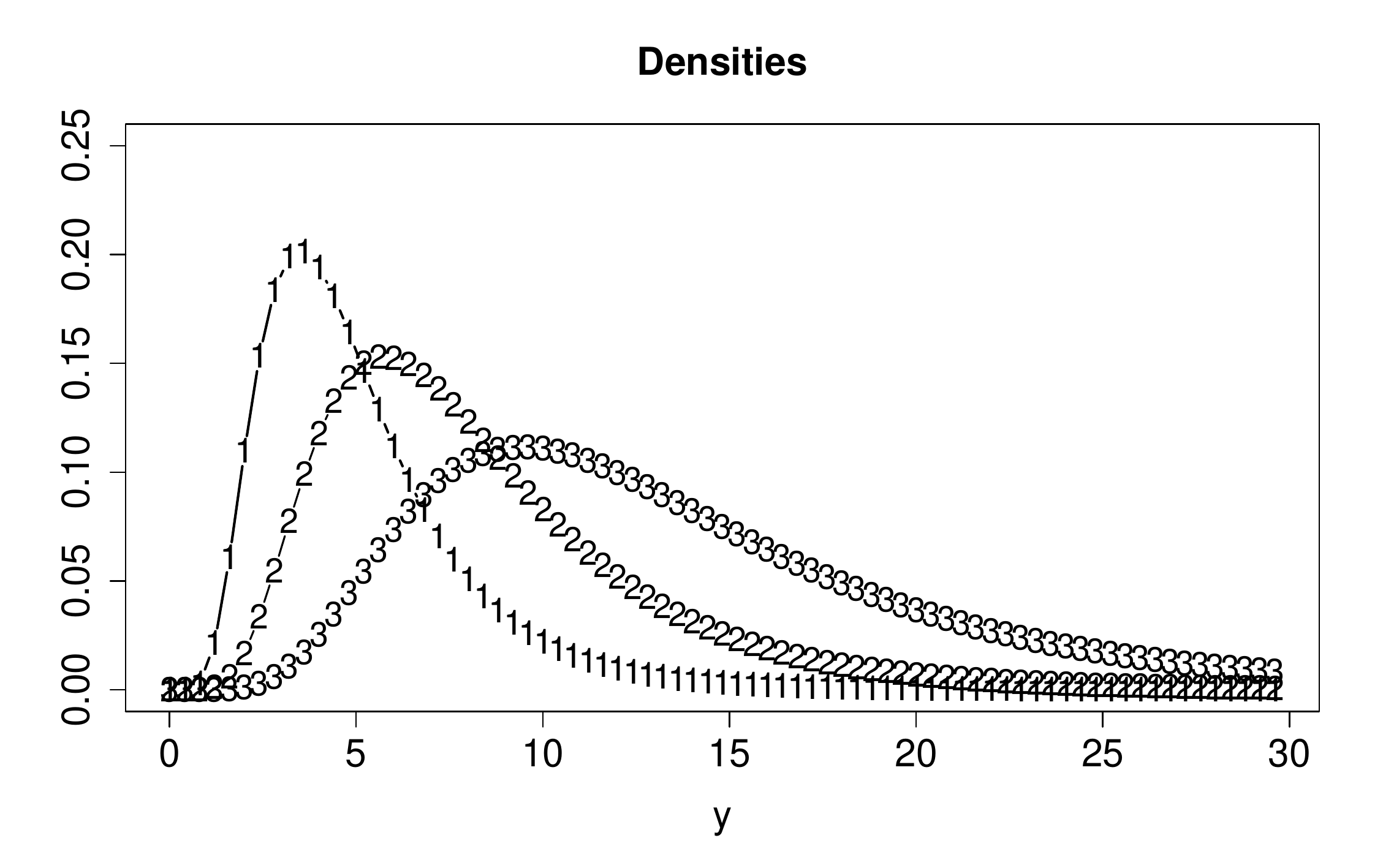}
\caption{Densities for three items, logarithmic difficulty functions, left: normal response function with intercepts, slopes given by $(-3,2), (-4,2), (-5,2)$, right:  Gumbel for $\bar F(.)$ with intercepts, slopes given by $(-3,2), (-4,2.5), (-5,3)$}
\label{fig:discr}
\end{figure}

For illustration Figure \ref{fig:discr} shows the response distributions for three items if the difficulty function is the logarithmic function. The left picture shows the distribution if the response function is the normal distribution, on the right side the Gompertz distribution has been used as response function.

Expectations and variances of responses are no longer linear functions of the person parameter. For the logarithmic function one obtains
\[
E(Y_{pi})= c_i\exp\left(\frac{\theta_p-\delta_{0i}}{\delta_i}\right),
\]
where $c_i$ is a constant that depends on $\alpha_i,\delta_i$ (Proposition \ref{meanmoments}). Thus, expectations are exponential functions of the latent ability. The same holds for the central moments,
\[
\E(Y_{pi}-\mu_{pi})^k= c_{ii}\exp\left(k\frac{\theta_p-\delta_{0i}}{\delta_i}\right),
\]
where $c_{ii}$ is again a constant that depends on $\alpha_i,\delta_i$ (Proposition \ref{meanmoments}) and for the quantile function, which has the form 
\[
Q_{Y_{pi}}(q) = \exp((\theta_p - F^{-1}(1-q)/\alpha_i -\delta_{0i})/\delta_{i}).
\]

\subsubsection*{Decomposition}

One can again try to decompose into a true score and an error score by using the representation 
\[
Y_{pi}= \tau_{pi}+ \varepsilon_{pi},
\]
where $\tau_{pi}=\E(Y_{pi})$ and $\varepsilon_{pi}$ is implicitly defined by $\varepsilon_{pi}=Y_{pi}- \tau_{pi}$. However, the decomposition is quite different from the decomposition for linear difficulties in  (\ref{equ:decomp}) since now the distribution of the error score depends on $\theta_p$. Even the support of $\varepsilon_{pi}$ depends on $\theta_p$ since $\varepsilon_{pi} \ge -\tau_{pi}$.

The form of the expectation has the consequence that the total  score $Y_{p+}=\sum_i Y_{pi}$, which is often used to measure the ability, is
not appropriate. One has 
\[
\E(Y_{p+}) = \sum_{i=1}^I c_i\exp\left(\frac{\theta_p-\delta_{0i}}{\delta_i}\right)= \sum_{i=1}^I c_i \exp(-\delta_{0i}/\delta_i)\exp(\theta_p)^{1/\delta_i},
\]
which is a weighted sum of exponential terms with the terms depending on the item (Proposition \ref{meanmoments}). Conditions under which the total score is an appropriate measure have been investigated in particular for catgorical responses, see, for example,  \citet{Masters:82,hemker1997stochastic,sijtsma2000taxonomy,hemker2001measurement}. In the present case a condition is that items are homogeneous, that is, $\delta_i=\delta_i$ for all $i$. In this case one obtains 
\[
\E(Y_{p+}) = \sum_{i=1}^I c_i \exp(-\delta_{0i}/\delta)\exp(\theta_p/\delta)= c \exp(\tilde\theta_p),
\]
where $\tilde\theta_p=\theta_p/\delta$ is the scaled ability and $c= \sum_{i} c_i \exp(-\delta_{0i}/\delta)$. Thus, in the homogeneous case $\E(Y_{p+})$ is an exponential function of the scaled ability $\tilde\theta_p$. Of course, one can also consider the transformed ability  $\exp(\tilde\theta_p)$ as a measure of ability. Then, $\E(Y_{p+})$ depends linearly on the (transformed) ability. Thus, in the homogeneous  case, but only then, the total score can be considered as representing the underlying ability.

\subsubsection*{Linearity}
It is interesting that at the heart of all thresholds models there is a linear relationship between abilities and expectations, however it does not relate to the expectations of the responses itself but to transformed responses. More concise, one can derive the general result 
\begin{align}\label{eq:lin2}
&\E(\delta_{i}(Y_{pi})) = \theta_p -  \mu_F/\alpha_i.
\end{align}
where $\mu_F,\var_F$ are again the expectation and variance corresponding to distribution function $F(.)$. For the variances one obtains $\var(\delta_{i}(Y_{pi})) = \var_F/ (\alpha_i\delta_i)^2$ (Proposition \ref{try}).

One consequence is that the expected \textit{transformed} total score $Y_{p+}^{(\delta)}=\sum_i \delta_{i}(Y_{pi})$ is a linear function of the ability,
\[
\E (Y_{p+}^{(\delta)})=  I \theta_p -  \sum_{i=1}^I\mu_F/\alpha_i.
\]
Thus,  $Y_{p+}^{(\delta)}$ can be seen as an indicator of the ability.  It has the form  
\[
Y_{p+}^{(\delta)}=   \sum_{i=1}^I  \delta_{0i}+ \sum_{i=1}^I\delta_i g(Y_{pi}),
\]                                
which is a weighted sum of transformed responses. This suggests that one could also work with the transformed responses  $g(Y_{pi})$ and formulate models for the transformed responses. Let us consider  the thresholds model with transformation $g(.)$, TM($F$, \{$\delta_{0i}+ \delta_i g(y)\}$), which is given by
\[
P(Y_{pi} > y|\theta_p,\alpha_i,\delta_{i}(.))=F(\alpha_i(\theta_p-\delta_{0i}- \delta_i g(y))).
\]
If the model holds one obtains for the transformed responses $g(Y_{pi})$  
\[
P(g(Y_{pi}) > z|\theta_p,\alpha_i,\delta_{i}(.))=P(Y_{pi} > g^{-1}(z) |\theta_p,\alpha_i,\delta_{i}(.))=F(\alpha_i(\theta_p-\delta_{0i}- \delta_i z)),
\]                                                        
which means that $g(Y_{pi})$ follows the  \textit{linear} threshold model TM($F$, \{$\delta_{0i}+ \delta_i y$\}) with the same item parameters.
Thus, one can alternatively fit the corresponding linear model for the transformed responses.  

More generally, one can establish a simple relationship between the TM and the chosen response function $F(.)$, if the difficulty function is fixed. It can be shown (see Proposition \ref{transform_delta} in the appendix) that    
\[
\delta(Y_{pi}) \text{  has the same distribution as  } \theta - {Y_0}/{\alpha_i},
\]
where $Y_0$ follows the distribution function $F(.)$
Hence, on the level of the transformed variable, the latent ability $\theta$ acts as a simple location parameter, thereby shifting expectations and quantiles in a linear manner, as already  described previously.

However, one has to be careful when comparing alternative models since the log-likelihoods of TM($F$, \{$\delta_{0i}+ \delta_i g(y)$\}) for the original data and TM($F$, \{$\delta_{0i}+ \delta_i y$\}) for the transformed data $g(y_i)$ are not the same. As an example we use the self-regulation data to be considered later (Section \ref{appl:self}). The log-likelihood obtained when fitting a model with normal response function and logarithmic difficulty function is -654.136, when fitting a model with linear difficulty function to the log-transformed data one obtains 326.531.  These models should definitely not be compared via goodness-of-fit measures based on their loglikelihoods although parameter estimates for both models are identical.

\subsubsection*{Special Cases}
If one assumes for $F(.)$ the standard normal distribution and  a logarithmic difficulty function the density of responses becomes 
\begin{align*}
f_{pi}(y)&=f(\alpha_i(\delta_{0i}+ \delta_i \log(y)-\theta_p))\alpha_i\delta_i/y\\
&= (\sqrt{2\pi})^{-1}\exp(-(\alpha_i(\delta_{0i}+ \delta_i \log(y)-\theta_p))^2/2)\alpha_i\delta_i/y \\
&= \frac{1}{y\sqrt{2\pi}\bar\sigma_i}\exp(-\frac{(\log(y)-\bar\mu_{pi})^2}{2\bar\sigma_i^2}),
\end{align*}
where  $\bar\mu_{pi}=(\theta_p-\delta_{0i})/\delta_i$, $\bar\sigma_i=1/(\alpha_i\delta_i)$. This is the lognormal distribution with parameters 
$\bar\mu_{pi}, \bar\sigma_i$. The homogeneous version of the model ($\delta_i=\delta$)  is equivalent to van der Linden's lognormal response-time model, which is a speed model that carefully distinguishes between time and speed \citep{van2016lognormal}.

The thresholds version of van der Linden's  model is a generalization allowing for varying slope parameters. It also offers the possibility to consider alternative response functions that replace the normal distribution and might yield better fit (see the application in Section \ref{appl:rot}).

\subsection{Responses in an Interval}

Let again the response function $F(.)$ be chosen as fixed. If responses are known to be restricted to the interval $(a,b)$, then for the difficulty function 
$\lim_{y\rightarrow a}\delta_i(y)=-\infty$ and $\lim_{y\rightarrow b}\delta_i(y)=\infty$ should hold.
A candidate  is the logit function $g(y)=\log(y-a)/(b-y)$ yielding $\delta_{i}(y)= \delta_{0i}+ \delta_i \log(y-a)/(b-y)$.
For simplicity one can also transform the data into the interval $(0,1)$ and use the simpler function $g(y)=\log(y)/(1-y)$.

\begin{figure}[H]
\centering
\includegraphics[width=7cm]{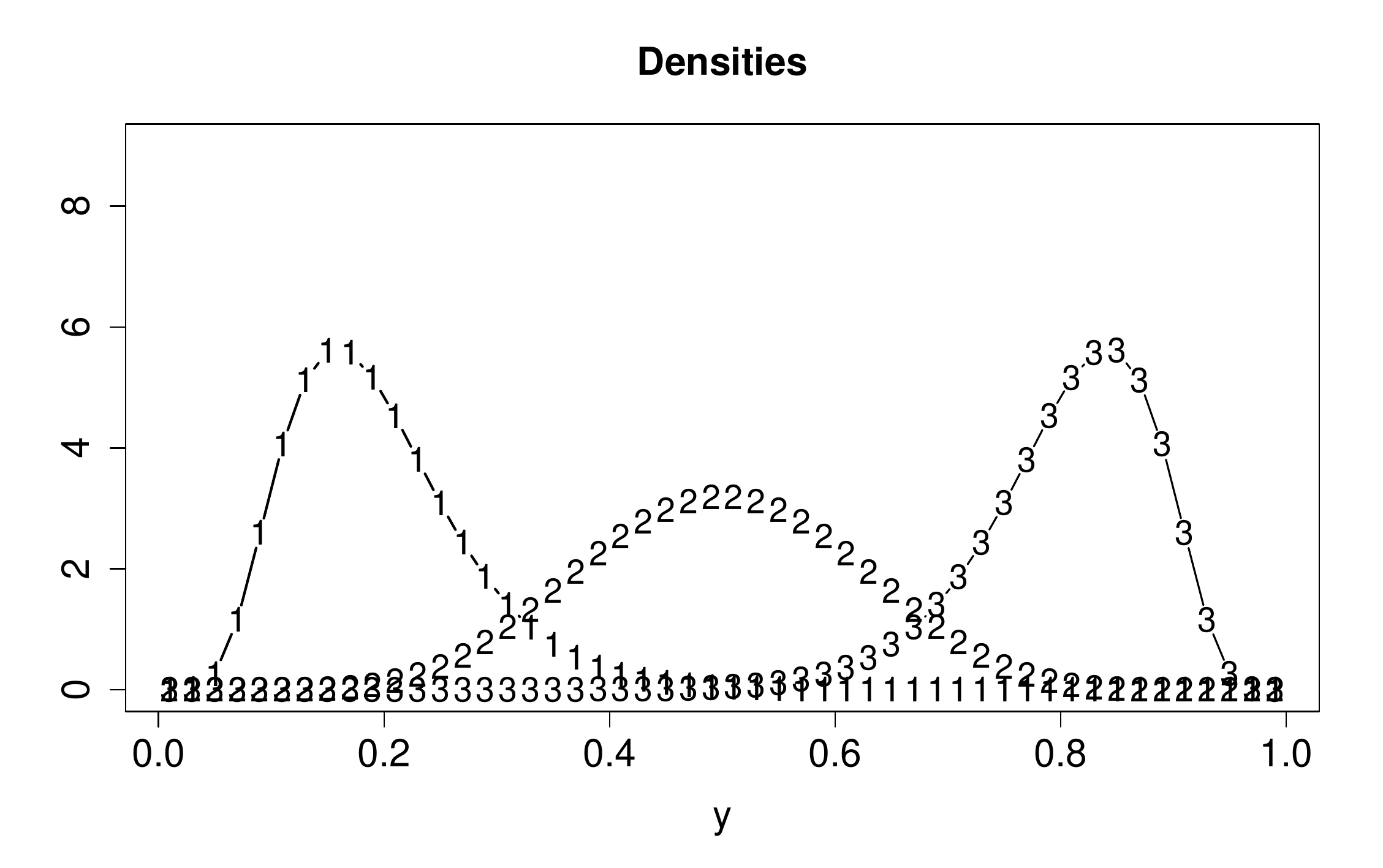}
\includegraphics[width=7cm]{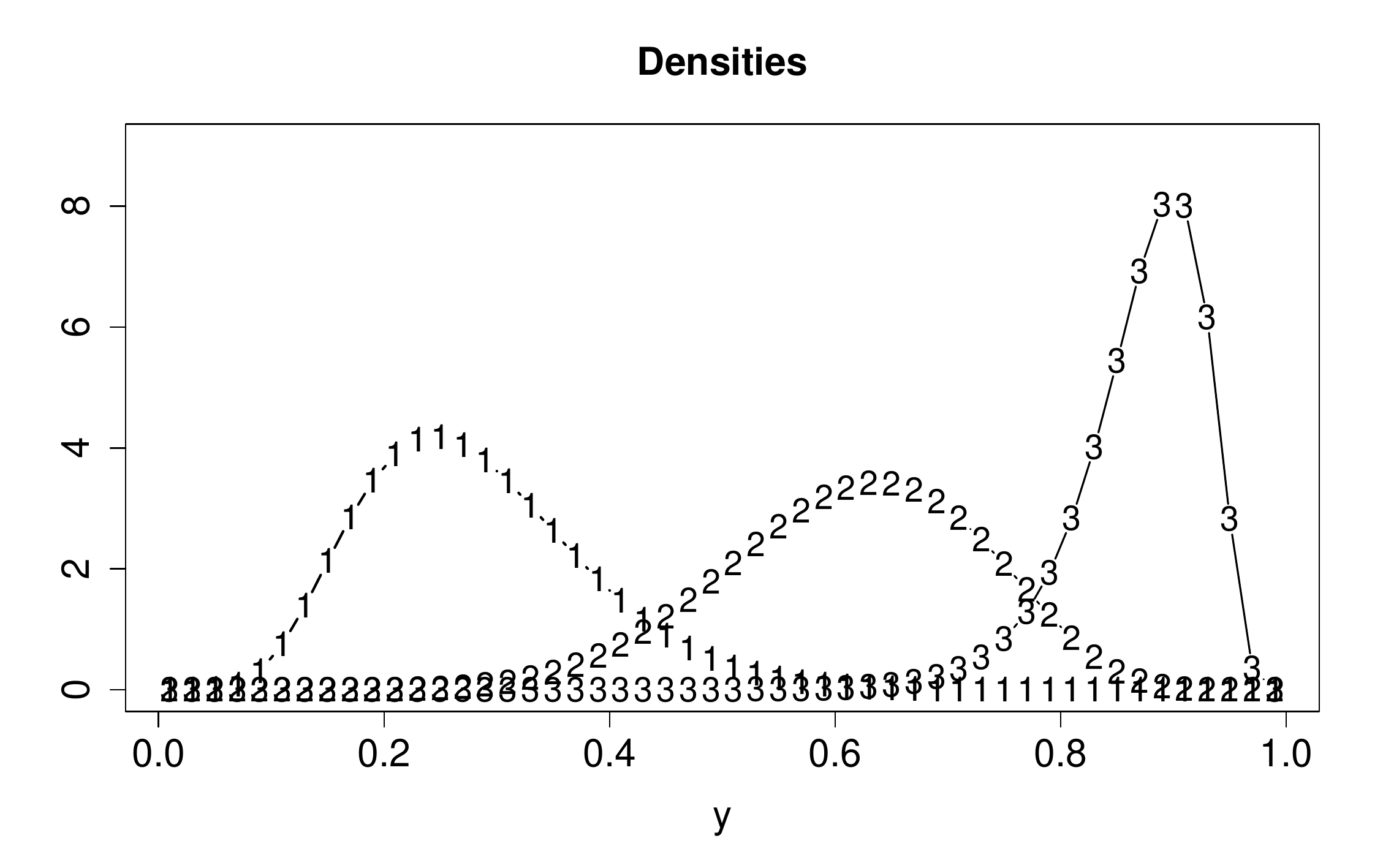}
\caption{Distributions of responses for items with intercept, slope given by $(3,2), (0,2), (-3,2)$ with logit difficulty function, left: $\theta=0$, right: $\theta=1$. }
\label{fig:intill}
\end{figure}

For illustration Figure \ref{fig:intill} shows the distributions that are obtained for the interval $(0,1)$ and standard normal distribution $F(.)$. The underlying item parameters are
$(3,2), (0,2), (-3,2)$ (for $(\delta_{0i},\delta_{i})$). The left picture shows the distribution of responses if $\theta_p=0$, in the right picture 
$\theta_p=1$. It is seen that responses are within the interval  $(0,1)$. For larger values of $\theta_p$ the distribution is not just shifted but distinctly changes its form.

For the expectation and variances of responses one obtains rather complicated formulae, which are not given. As in the case of logarithmic difficulty functions they depend on the person's ability $\theta_p$. Also the simple total score depends on the ability in a complex form and can not be considered an appropriate measure of the underlying latent trait. However, one can use again the transformed total score $Y_{p+}^{(\delta)}=\sum_i \delta_{i}(Y_{pi})$, for which the relationship
\[
\E (Y_{p+}^{(\delta)})=  I \theta_p -  \sum_i\mu_F/\alpha_i  
\]
holds (following from the general result (\ref{eq:lin2}). It is noteworthy that the transformed total score is quite different from the traditional total score since the transformation is highly non-linear. However, it is much more appropriate as an indicator of the underlying ability than the traditional score given by sums of responses.

The case of responses in intervals is especially important in Likert-type items, which by definition are restricted to a fixed interval $[1,m]$ in $m$-grade 
Likert scales. As already shown in Figure \ref{fig:curves1} ignoring the restriction to an interval yields improper densities, which typically are positive beyond the interval $[1,m]$. In addition, the total score as a sum of responses over items is not a reliable indicator of the latent trait. 

\section{Applications}

We   illustrate the usage of the nonlinear TM models with three examples. Along with the modeling of properly continuous data, like response times, we will   in particular focus on the practical usage of the application of these models to Likert scales, whereby in contrast to a direct linear, unrestricted treatment, we take care of the range of the restricted support via appropriately chosen difficulty functions. 
 
All models are fitted using the MML-procedure and Gauss-Hermite quadrature, whereby a centered normal distribution with unknown variance $\sigma^2_{\theta}$ for the latent variable is specified and wherein identifiability issues are resolved by fixing the item discrimination parameter on the first item to unity. The full R-Code is provided as supplementary material. We abstain from including the   likelihood and score functions, which  can be found in \citet{TuItThres2021}.

\subsection{Self-Regulation}\label{appl:self}

The data set \textit{Lakes} from the R package \textit{MPsychoR} \citep{mair2018modern}  is a multi-facet G-theory application  taken from \citet{lakes2009applications}. The authors used the  response to assess children's self-regulation in
response to a physically challenging situation. The scale consists of three domains, cognitive, affective/motivational, and physical. We use the physical domain only. Each of the 194 children was  rated by 5 raters on six items on his/her self-regulatory ability with  
ratings on a scale from 1 to 7. We use the average rating over the five raters, which yields a response that takes values in the interval $(1,7)$ but is not confined to integer values. 
 
Table \ref{tab:phys } shows log-likelihoods and estimates of $\sigma_{\theta}$ for various response and difficulty functions with varying slopes in the difficulty functions.  The columns on the left show fits for fixed discrimination parameters ($\alpha_i=1$), the right columns show fits for varying discrimination parameters. It is seen that varying discrimination yield significantly better fits. For example, the likelihood ratio test that compares the fixed discrimination model and the varying discrimination model is 19.468 on 2 df for the Gumbel model with logit difficulty function. Similar results hold for the other models.   For  given response function $F$, the best fits are always obtained via logit difficulty functions.   Conversely, for a given difficulty function, the best fit is obtained by choosing the Gumbel distribution as   response function $F$. Note that the   best fit is indicated by an asterisk.

\begin{table}[H]
 \caption{Fits for self-regulation data } \label{tab:phys }
\centering
\begin{tabularsmall}{llllllllcccccccccc}
  \toprule
 &\multicolumn{2}{c}{ } &\multicolumn{2}{c}{ $\alpha_i=1$ } &\multicolumn{2}{c}{$\alpha_i $ varying}\\
 \midrule
 &response fct  & difficulty fct & log-likelihood & $\hat\sigma_{\theta_p}$ & log-likelihood & $\hat\sigma_{\theta_p}$\\    
  \midrule
&NV&lin & -563.084 &1.493 &-554.421   &2.596\\
& & log &-654.136   &1.260 &-644.621   &2.379 \\
& & logit(c=0.10) &-533.616    & 1.660 &-516.132&2.572 \\
&Gumbel&lin & -525.132  &1.702  &-514.559   &2.643\\
&  &log & -552.629  &1.584 &-549.302 &2.607\\
& & logit(c=0.10) & -507.9695  & 1.957  &-497.235{\text{*}}  &2.925\\
&Gompertz&lin &-597.118    &1.837  &-594.366   &1.955\\
&  &log & -694.613   &2.174 &-684.997   &2.818\\
& & logit(c=0.10) & -563.996    &1.796  &-554.913   &1.900 \\
\bottomrule
\end{tabularsmall}
\end{table}

\begin{figure}[H]
\centering
\includegraphics[width=7cm]{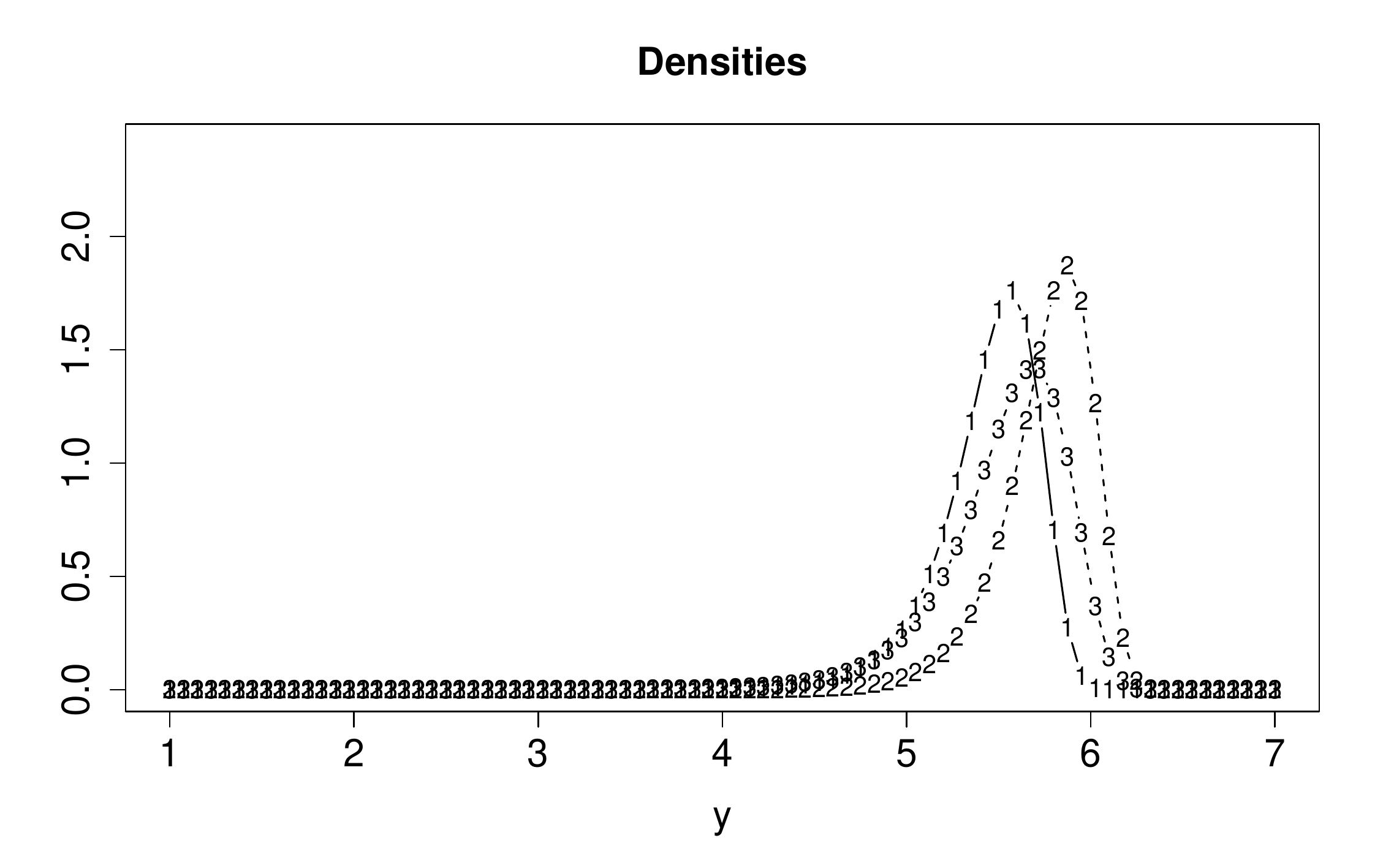}
\includegraphics[width=7cm]{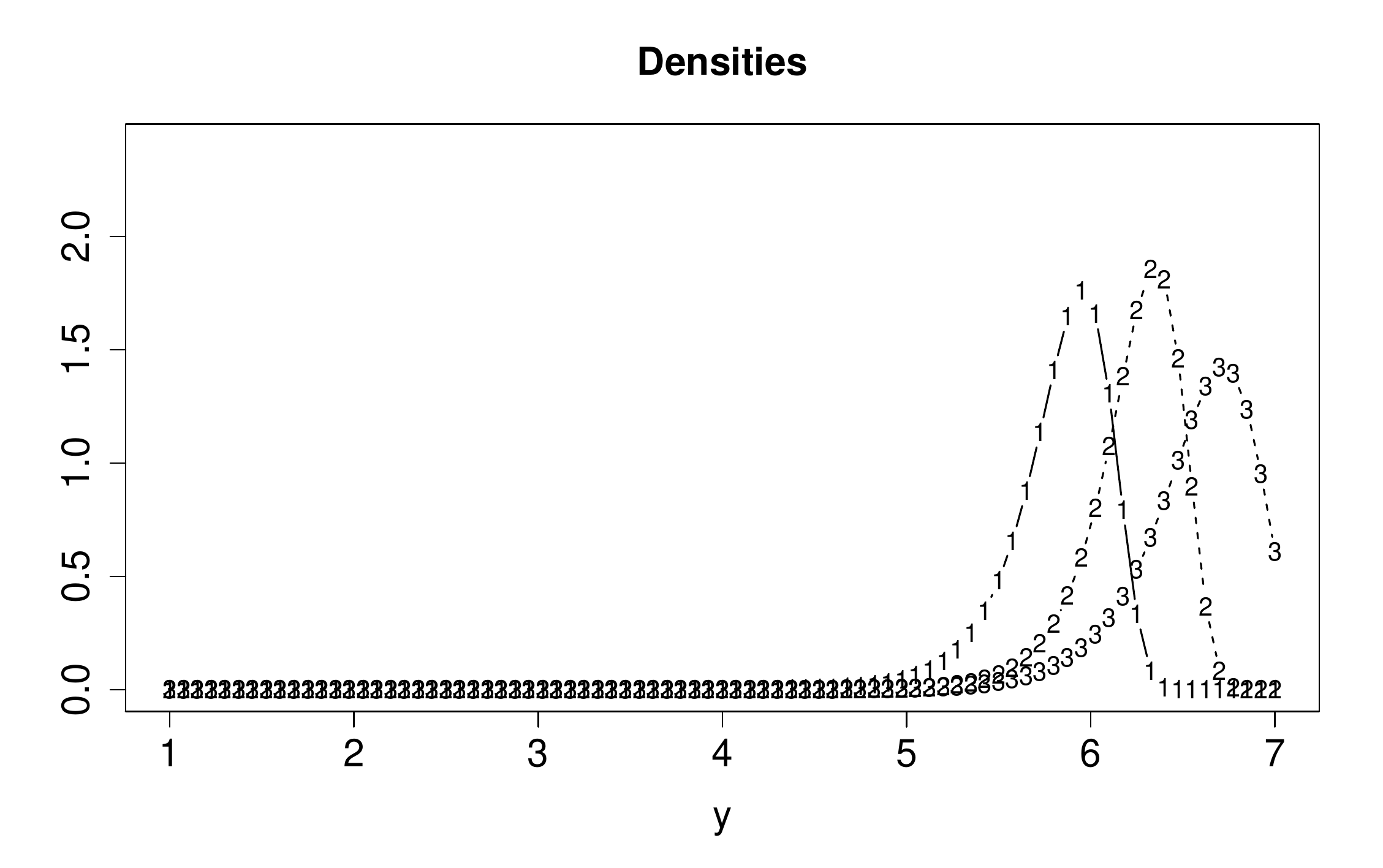}
\includegraphics[width=7cm]{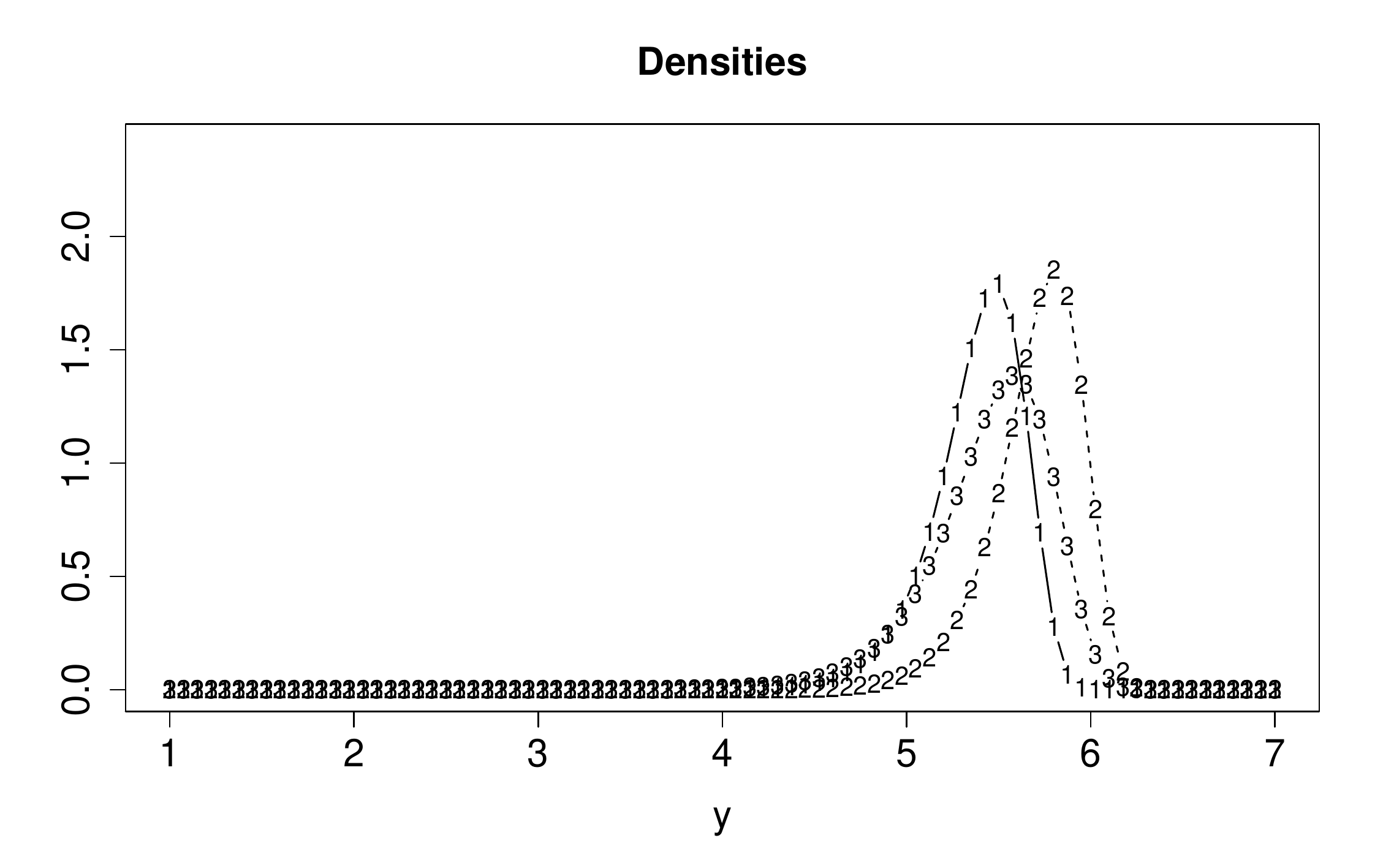}
\includegraphics[width=7cm]{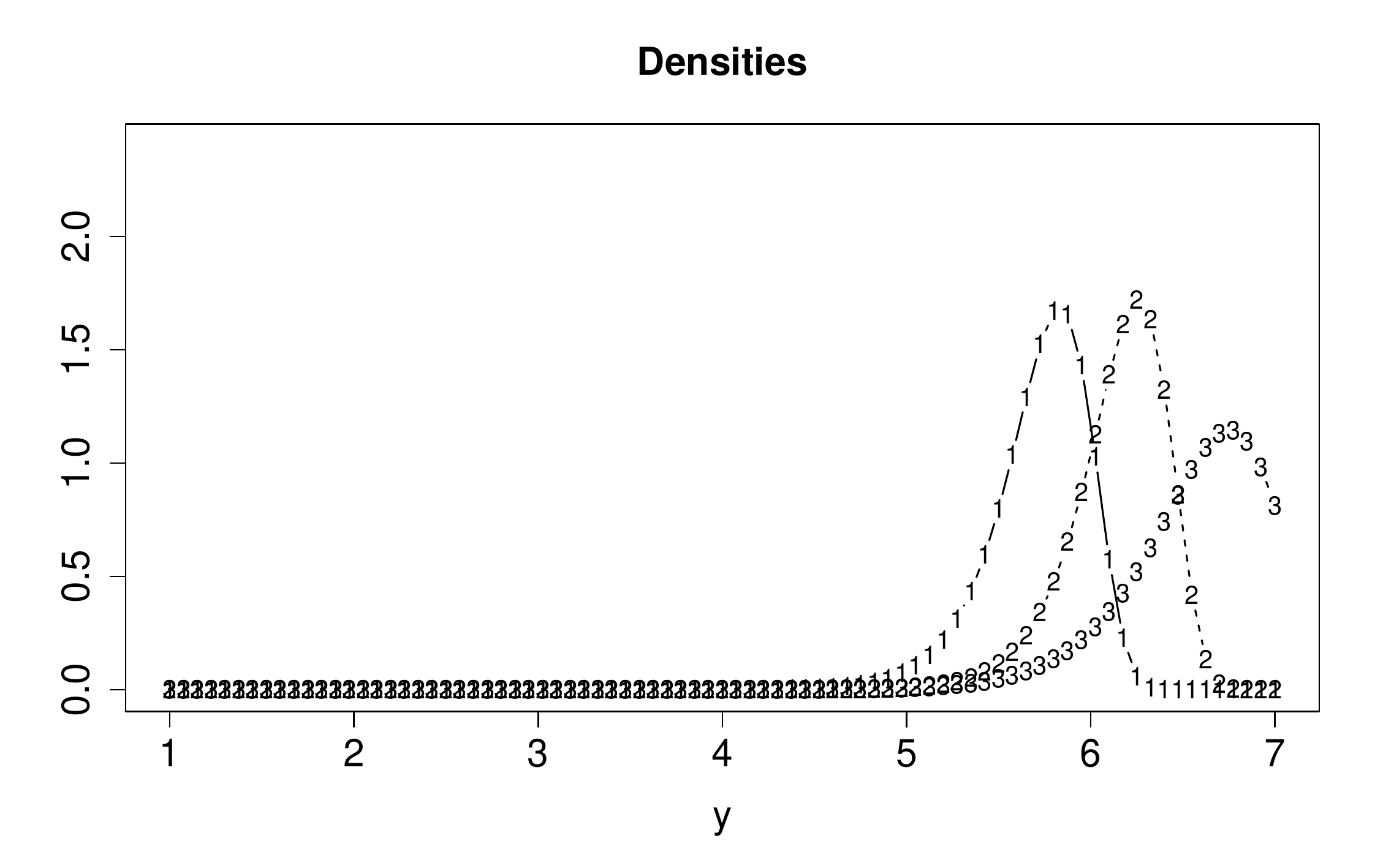}
\includegraphics[width=7cm]{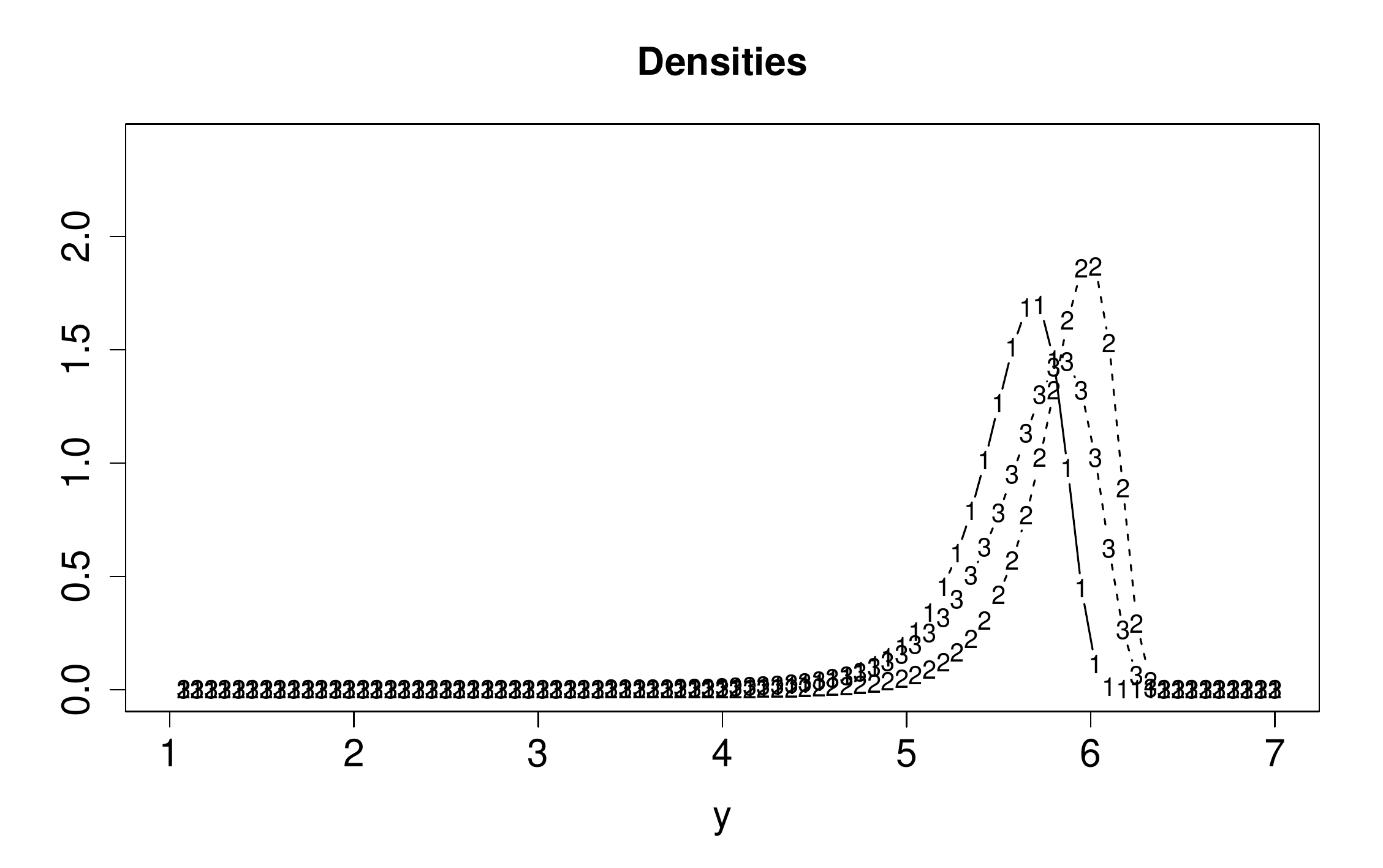}
\includegraphics[width=7cm]{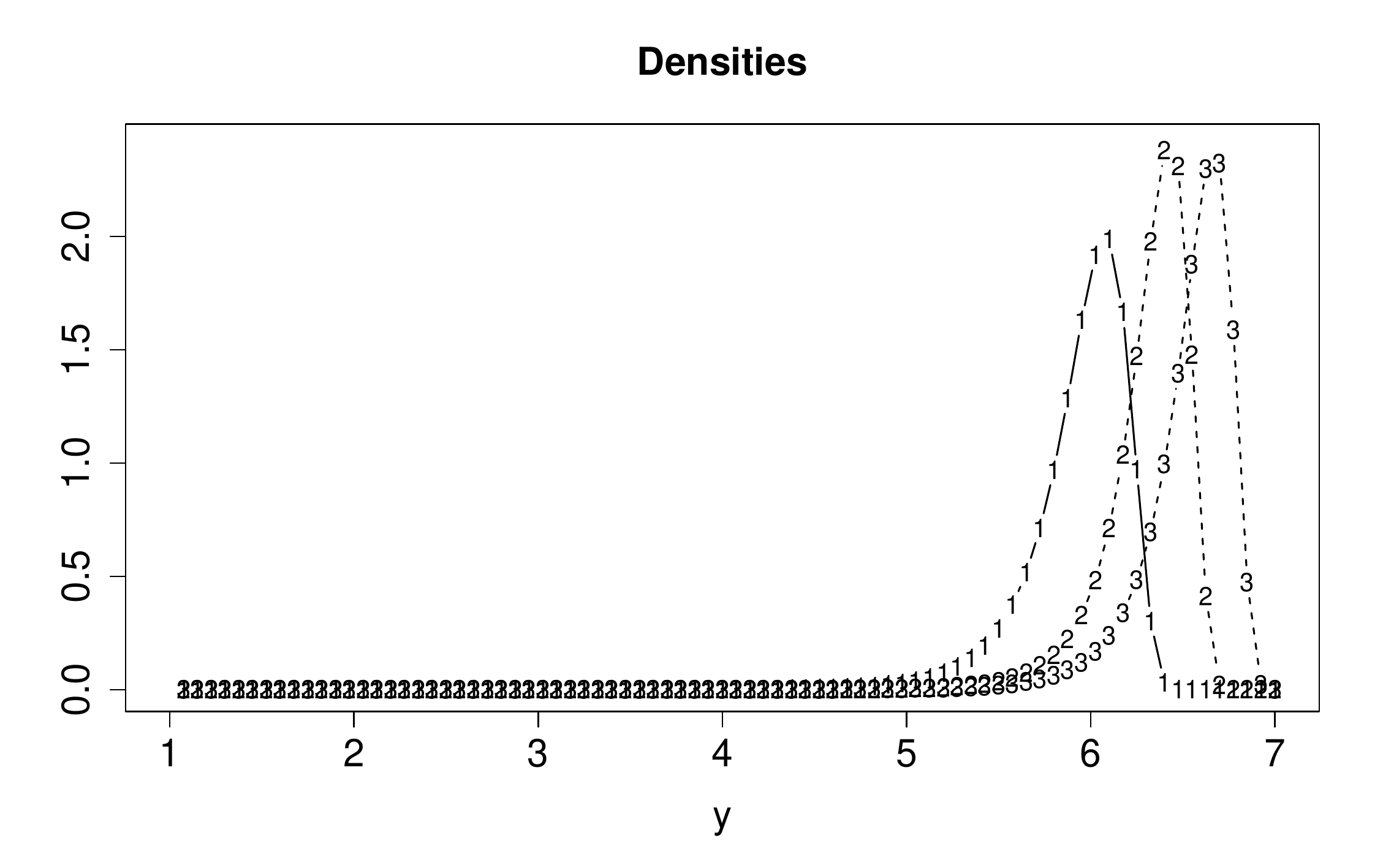}
\caption{Estimated densities for self-regulation,  first row: linear difficulty functions,   second row: logarithmic difficulty functions,  third row: logit difficulty functions, left: $\theta_{\text{low}}=0$, right: $\theta_{\text{high}}=1.5$). }
\label{fig:selfill}
\end{figure}

Figure \ref{fig:selfill} shows the estimated response densities for the three items with varying discrimination parameters (using as response function the Gumbel distribution). First row shows linear difficulty functions,   second row   logarithmic difficulty functions, and third row  logit difficulty functions, left column shows responses for latent trait $\theta_{\text{low}}=0$, right column shows responses for latent trait $\theta_{\text{high}}=4$, which is not extreme given $\hat\sigma_{\theta_p}$  is larger than 2.5.
It is seen that linear and logistic difficulty functions yield inappropriate densities that take values outside the interval $(1,7)$. The logit difficulty function, which restricts the responses, yields much more appropriate distributions. 

\subsection{Rotation Response Time}\label{appl:rot}

The R package \textit{diffIRT}contains response time data
 of 121 subjects to 10  mental rotation items.  Each item consists of a graphical display of
two 3-dimensional objects. The second object was either a rotated version of the first object, or a
rotated version of a different object. Subjects were asked whether the second object was the same
as the first object (yes/no). The degree of rotation of the second object was either 50, 100, or 150
degrees.  Response times were recorded in seconds.

\begin{figure}[H]
\centering
\includegraphics[width=7cm]{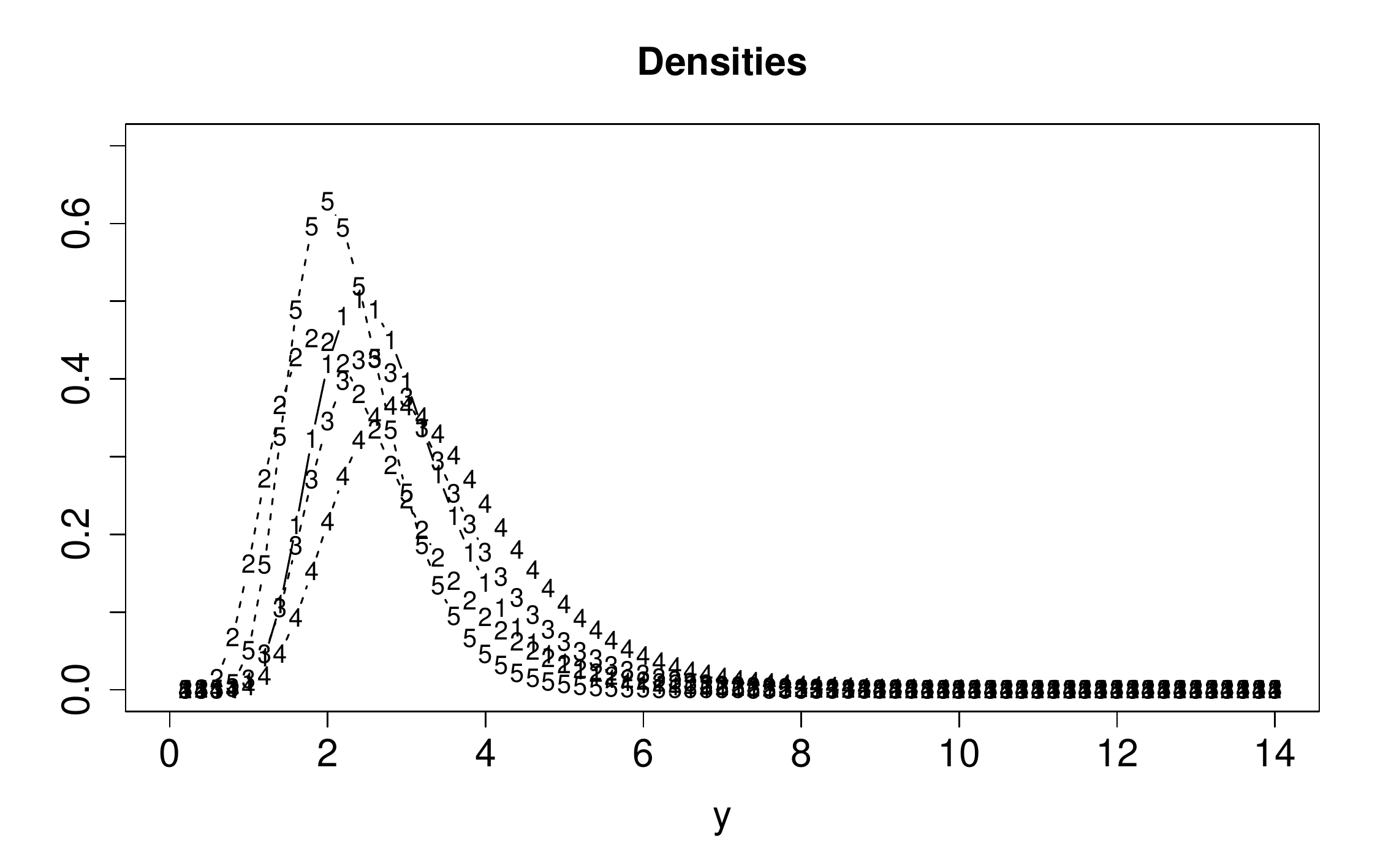}
\includegraphics[width=7cm]{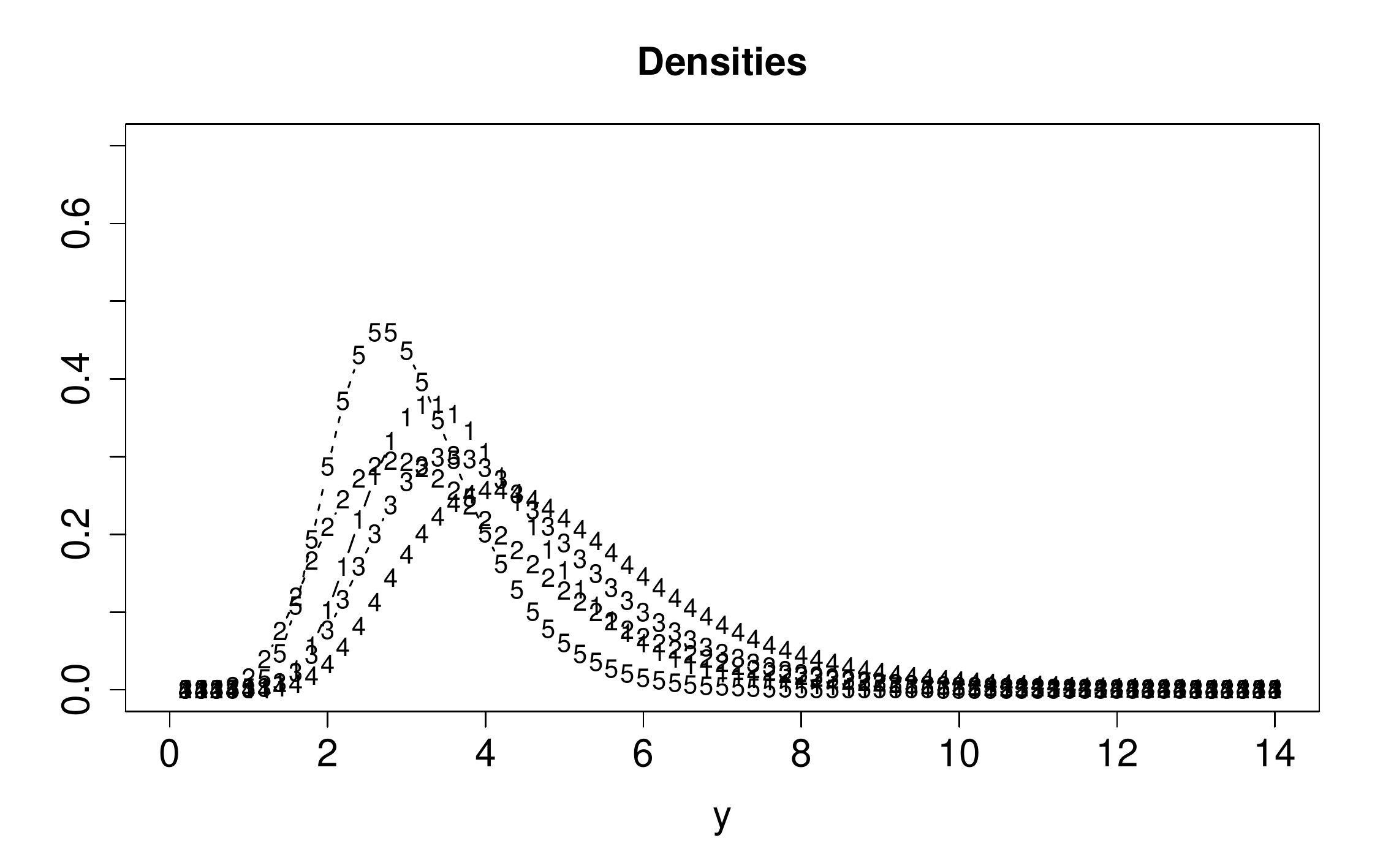}
\caption{Response distributions for the first five items of the mental rotation dataset for $\theta_p=0.0$ (left) and $\theta_p=1.0$ (right).}
\label{fig:rot1}
\end{figure}

We fitted  thresholds model with logarithmic difficulty function and fixed discrimination parameter. The best fit was obtained for normal response function (loglik
 -1300.378, $\sigma_{\theta}=0.817 $), which outperformed the Gumbel  response function  (loglik
 -1390.093) and the Gompertz response function  (loglik -1321.674). Testing if slopes in the difficulty functions can be modeled as constant yields the likelihood ratio statistic
 32.486 on 9 df, which indicates that slopes  should be considered as varying across items (see also Table \ref{tab:respt1}). Thus the simple lognormal model proposed by \citet{van2016lognormal} seems inadequate. The fit of the model could additionally improved by allowing for varying discrimination parameter. The corresponding log-likelihood was -1296.027, however it does not significantly improve the fit (log-likelihood test is 8.702 on 9 df).
 Figure \ref{fig:rot1} shows the estimated response distributions for the first five items  for $\theta_p=0.0$ (left) and $\theta_p=1.0$ (right) for the normal response function and varying slopes.

\begin{table}[H]
 \caption{Parameter estimates for response time data with varying and constant slopes} \label{tab:respt1}
\centering
\begin{tabularsmall}{llllllllcccccccccc}
  \toprule
 &\multicolumn{1}{c}{ } &\multicolumn{2}{c}{ Varying slopes } &\multicolumn{2}{c}{ Common slopes }\\
 &item  & intercept & slope & intercept & slope\\    
  \midrule
 &[1] &-3.132322 &3.196110  &-2.873841 &2.950614\\
 &[2] &-1.866113 &2.316536  &-2.370173 &2.950614\\
 &[3] &-2.988711 &2.862885  &-3.067789 &2.950614\\
 &[4] &-3.336943 &2.817903  &-3.484121 &2.950614\\
 &[5] &-2.554667 &3.276066  &-2.289048 &2.950614\\
 &[6] &-2.995613 &3.628465  &-2.423716 &2.950614\\
 &[7] &-2.977589 &3.034445  &-2.883957 &2.950614\\
 &[8] &-1.917933 &3.012964  &-1.864209 &2.950614\\
 &[9] &-2.949433 &3.124232  &-2.774521 &2.950614\\  
&[10] &-3.699182 &2.991578  & -3.639867 &2.950614\\
\midrule
&loglik &-1300.378 &&-1316.621\\
\bottomrule
\end{tabularsmall}
\end{table}

\subsection{Political Fears}


We consider  data from the German Longitudinal Election Study (GLES), which is a long-term study of the German electoral process \citep{GLES}.  The data we are using  originate from the pre-election survey for the German federal  election  in  2017 and consist of responses to various items adressing political fears. The participants were asked: ``How afraid are you due to the ...'' - (1) refugee crisis?
- (2) global climate change?
- (3) international terrorism?
- (4) globalization?
- (5) use of nuclear energy?
The answers were measured on Likert scales from 1 (not afraid at all) to 7 (very afraid). The model is fitted under the assumption that fear is the dominating latent trait, which is considered as unidimensional. We  use 200 persons sampled randomly from the available set of observations. 

\begin{figure}[H]
\centering
\includegraphics[width=7cm]{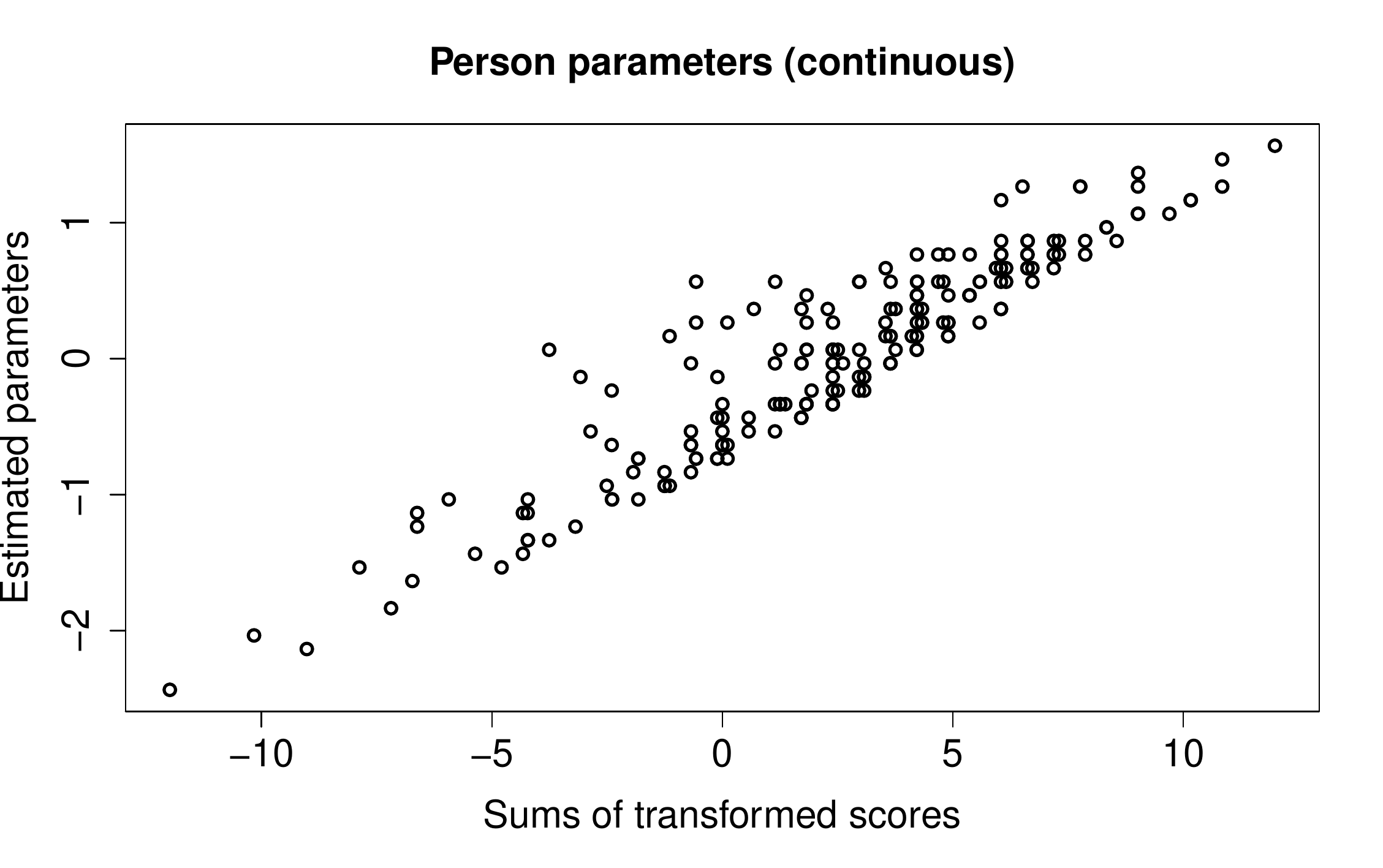}
\includegraphics[width=7cm]{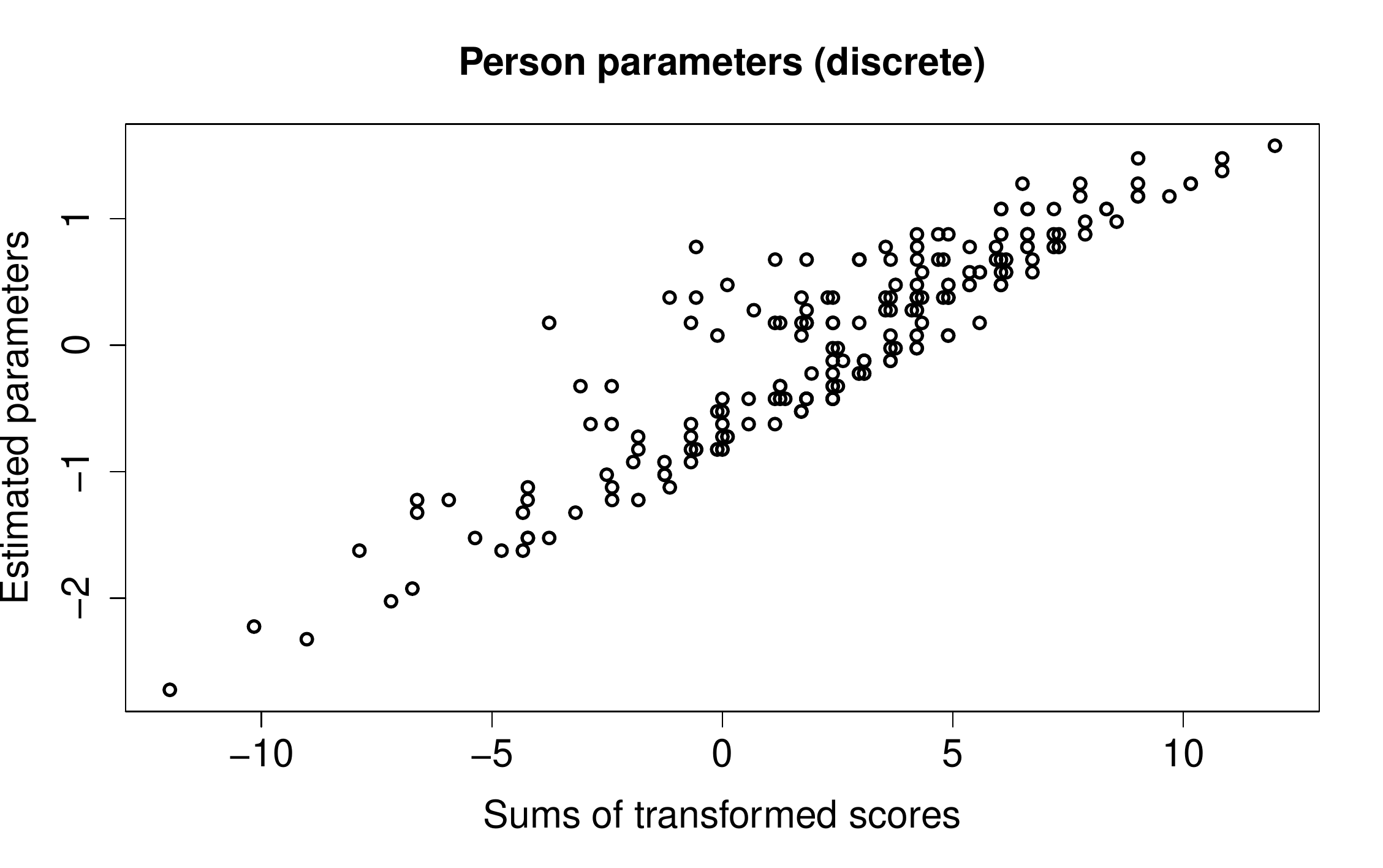}
\caption{Estimated person parameters for fears data. }
\label{fig:fearspers}
\end{figure}

Table \ref{tab:fear1} shows log-likelihoods and estimates of $\sigma_{\theta_p}$ for various responses and difficulty functions. The models have varying slopes in the difficulty functions and fixed discrimination parameters since varying discrimination parameters did not improve the fit significantly. We considered two fits,   assuming  that the responses are continuous or discrete. The latter means that one assumes a multinomial distribution, which changes the likelihood. It is seen that in both cases  the logit difficulty function always fits better than linear or logarithmic functions. The problems with the latter two have already been illustrated in  Figure \ref{fig:fearsill}. Again, the best fit was obtained by specifying a Gumbel response function, which was chosen as the best model under the assumption of continuously distributed responses  as well as under the assumption of discrete responses. For a further  investigation of the   difference between continuous and discrete modeling we computed the posterior estimates of person parameters. Figure \ref{fig:fearspers} shows the estimates plotted against the transformed sum scores $Y_{p+}^{(\delta)}$ using logit transformed data (difficulty function) and Gumbel response function. It is seen that the pictures for continuous and discrete responses are virtually the same. Correspondingly the correlation between estimated person parameters for discrete and continuous data was 0.995.  The correlation between transformed sum scores and estimated parameters was 0.926 in both cases. 
Thus,   concerning model selection as well as prediction of person parameters there seems to be no relevant difference between assuming a continuous distribution or a discrete distribution, although the concrete log-likelihoods differ for discrete and continuous distributions.

\begin{table}[H]
 \caption{Thresholds models for fears data} \label{tab:fear1}
\centering
\begin{tabularsmall}{llllllllcccccccccc}
  \toprule
 &\multicolumn{2}{c}{ } &\multicolumn{2}{c}{ Continuous } &\multicolumn{2}{c}{ Discrete }\\
 &response fct  & difficulty fct & log-likelihood & $\hat\sigma_{\theta_p}$ & log-likelihood & $\hat\sigma_{\theta_p}$\\    
  \midrule
&NV&lin & -1847.53 &0.707  &-1825.490  &0.746\\
& & log &-2046.301   &0.719 &-2064.723   &0.758 \\
& & logit(c=0.10) &-1714.907    &0.706  &-1802.259   &0.751 \\
&Gumbel&lin &-1817.634  &0.799 &-1781.461  &0.806\\
&  &log & -1891.852 &0.629  &-1915.859&0.619\\
& & logit(c=0.10) &-1713.209{\text{*}}   &0.850  &-1752.010{\text{*}} & 0.817\\
&Gompertz&lin &-1923.348  &0.885  & -1899.042 &0.969 \\
&  &log &-2178.687  &1.083 &-2179.356 &1.174\\
& & logit(c=0.10) &-1773.234   &0.915 &-1877.851   &1.035  \\
\bottomrule
\end{tabularsmall}
\end{table}

\section{Including Covariates}

The basic threshold model assumes that latent trait are unidimensional and not affected by covariates. If one suspects that covariates may modify the response behaviour it can be tested by including them explicitly in the explanatory term.  
Let $\xb_p$ be a person-specific vector of covariates. In  a threshold model with covariates the person parameter $\theta_p$ is replaced by $\theta_p-\xb_p^T\gammab_i$ yielding
\begin{align}\label{eq:dif}
P(Y_{pi} > y|\theta_p,\alpha_i,\delta_{i}(.))=F({\alpha_i(\theta_p-\xb_p^T\gammab_i-\delta_{i}(y))}),
\end{align}
where the parameter $\gammab_i$ is item-specific and represents the effect on the response in item $i$.

Within item response theory the inclusion of  covariates can be seen as investigating  
differential item functioning (DIF), which  is the well known phenomenon
that the probability of a correct response among equally able
persons may differ in subgroups, see, for example,
\citet{millsap1993methodology}, \citet{zumbo1999handbook}, \citet{rogers2005differential,osterlind2009differential},  \citet{magisetal:2010}. 
If $\gammab_i$ is non-zero the item functions differently in subgroups represented by covariates. 

Model (\ref{eq:dif}) can also be seen as a multivariate regression model with heterogeneity. If one is primarily interested in the effects of covariates one 
considers $\theta_p$ as representing the heterogeneity needed to model the effects adequately. It can be seen  as a generalized random effects model, but  
with much weaker assumptions on the disribution of the response variables than in the classical linear mixed model \citep{Goldstein:87,SeaCasMcC:92}.

Let us consider the fear data with covariates 
gender (1: female; 0: male) and age in years.
Table \ref{tab:fearscov1} shows the estimates for the basic model without covariates and the model with covariates age and gender (continuous response, $\alpha_i=1$).
It is seen that   all items show significant covariate effects for at least one of the covariates ( z-values given in the last two columns).  With the exception of climate change  older respondents tend to be more afraid than younger respondents, females have for all items higher fear levels than males.
The necessity of covariates is also supported by testing. The log-likelihood test that compares the model without covariates to the model with covariates is 41.266 on 10 df. Thus, if one considers it as a DIF problem, all items show differential item functioning. From a regression perspective, gender and age are seen to be influential if one accounts for the heterogeneity in the population. 

\begin{table}[H]
 \caption{Parameter estimates for the fears data with logit difficulty function, Gumbel distribution response function without covariate and with covariates, the last two columns show the z-values of parameter estimates of covariate parameters. } \label{tab:fearscov1}
\centering
\begin{tabularsmall}{lllrrrrrrcccccccccc}
  \toprule
 &\multicolumn{2}{c}{ } &\multicolumn{4}{c}{ Parameters } &\multicolumn{2}{c}{ z-values }\\
 \midrule
 &&Item  & intercepts & slopes & Age & Gender & z-Age & z-Gender\\ 
  \midrule
&1 &&-0.751  &1.0259 \\
&2 & &-1.459  &1.2609 \\
&3 & &-2.134  &1.3365 \\
&4 & &-0.360  &1.2061 \\
&5 & &-1.125  &1.1303 \\
\midrule
&&Log-lik &-1713.209 &&\\
\midrule
&1&refugee& 0.072    &1.028 &-0.011   &-0.487 &-2.138  &-2.548 \\
&2&climate change& -1.055   &1.262 &-0.002 &-0.642  &-0.326  &-3.328 \\
&3&terrorism& -0.954  &1.393 &-0.017   &-0.756  &-3.379  &-3.997 \\                  
&4&globalization& 0.416   &1.201 &-0.010   &-0.491  &-2.085  &-2.603 \\
&5&nuclear energy& -0.385   &1.131 &-0.013  &-0.089    &-2.632  &-0.474 \\
\midrule
&&Log-lik &-1692.576 &&\\
\bottomrule
\end{tabularsmall}
\end{table}

\section{Concluding Remarks}
The topic of latent trait modeling for continuous responses has been addressed within the framework of thresholds models.  With respect to continuous responses, the lognormal and the normal-linear model (the basic building block in the factor analysis model) have been shown to be members of the thresholds modeling class. Furthermore, a better approximation to the handling of Likert-type data has been suggested via the usage of appropriately chosen nonlinear difficulty functions. It has also been demonstrated that response functions other than the normal distribution can be more appropriate.  

Future research could focus on multidimensional extensions of the TM class which would ultimately provide a latent trait model for multidimensional abilities and continuous responses. Alongside these multidimensional extensions, the modeling of data for mixed measurement levels also becomes important. For instance, response times are usually recorded in conjunction with the accuracy of the response (correct/incorrect). A proper approach would need to model the joint distribution of $(x_i,t_i)$ (with $x_i$ denoting the binary indicator of a correct response) in terms of a multdimensional trait encompassing a speed and an accuracy component.
Finally, it should be emphasized that for continuous responses, the sensitivity of parameter estimates with respect to extreme responses becomes of extra importance. That is, when using these models for the classification of persons, it has to be ruled out that extreme responses on single items show large effects  on the estimate of the ability parameter. The answer to this question might depend on the choice of the difficulty function and the choice of the response function in the TM model and is an additional topic of future research. 


%
%
%
%
%

\bibliography{literatur_PJ}

\newpage
\section*{Appendix}

\subsection{Classical test theory}

In the following we show that given a general TM model holds,    by appropriate definitions of the true score and error terms 
a CTT model can be shown to hold, see also \citet{holland2003classical}.

\begin{theorem}\label{ctt_theorem}
Define a true score by $T_i(\theta):=E(Y_i|\theta)$ and an error score via $\varepsilon_i(\theta):=Y_i-T_i$.
Then the following holds for all choices of $i$ and $j$:
\begin{itemize}
\item[a)] $\mathbb{E}(\varepsilon_j|T_i) = 0$, 
\item[b)] $\mathbb{E}(\varepsilon_j) = 0$,
\item[c)] $\text{Cov}(\varepsilon_i,\varepsilon_j) = 0$ for $i\neq j$,
\item[d)] $\text{Cov}(\varepsilon_i,T_j) = 0$,
\item[e)] $\text{Cov}(Y_i,Y_j) = \text{Cov}(T_i,T_j) $ for $i\neq j$.
\end{itemize}
\end{theorem}
\begin{proof}
First note that we may in the following occasionally drop the dependency of $T$ and of $\varepsilon$ on $\theta$ and simply write $T$ and $\varepsilon$ instead of $T(\theta), \varepsilon(\theta)$. 

$(a)$ Due to the strict monotonicity of the function $T_j(\theta):=E(Y_j|\theta)$, $T_j$ and $\theta$ are in a one-to-one relationship. Therefore, we have
$$\mathbb{E}(\varepsilon_j|T_i) = \mathbb{E}(\varepsilon_j|\theta)=\mathbb{E}(Y_j-T_j(\theta)|\theta)=\mathbb{E}(Y_j|\theta)-T_j=0$$. 

$(b)$ A direct consequence of (a) and the law of iterated expectations.

$(c)$ Due to $(b)$ it suffices to show that $\mathbb{E}(\varepsilon_i \varepsilon_j)=0$.
\begin{align*}
\mathbb{E}(\varepsilon_i \varepsilon_j) =&\mathbb{E}\left((Y_i-T_i) (Y_j-T_j)\right)=\mathbb{E}\left(\mathbb{E}\left((Y_i-T_i) (Y_j-T_j)|\theta\right)\right) 
 \\
=&\mathbb{E}\left(\mathbb{E}(Y_i-T_i|\theta)\mathbb{E}(Y_j-T_j|\theta) \right) = \mathbb{E}(0 \cdot 0) = 0.
\end{align*}
In the above factorization we used the assumption of local stochastic independence.

$(d)$ We have
\begin{align*}
\text{Cov}(\varepsilon_i,T_j) =& \text{Cov}\left(\mathbb{E}(\varepsilon_i|\theta),\mathbb{E}(T_j|\theta)\right) + \mathbb{E}\left(\text{Cov}(\varepsilon_i,T_j|\theta)\right).
\end{align*}
As $T_j$ is constant given $\theta$, the second term may be dropped. Likewise, since $\mathbb{E}(\varepsilon_i|\theta)=0$ according to (a) the first term may also be dropped.

$(e)$ \begin{align*}
\text{Cov}(Y_i,Y_j) =& \text{Cov}(T_i+\varepsilon_i,T_j+\varepsilon_j) \\
=& \text{Cov}(T_i,T_j) + \text{Cov}(T_i,\varepsilon_j) + \text{Cov}(\varepsilon_i,T_j) + \text{Cov}(\varepsilon_i,\varepsilon_j)
\end{align*}
The last three terms cancel due to $(c)$ and $(d)$.
\end{proof}

\subsection{Quantile function and moments}

Equation (\ref{eq:quant}) provides us with a direct formula for the quantile function.
Based on this, we can use the following result (with slight changes adapted from Theorem 11a in \citet{widder2015laplace}) -- which we provide here for completeness -- to compute expectations of functions of $Y_{pi}$.
\begin{theorem}
Let $h, F$ be functions from $[a,b] \mapsto \mathbb{R}$. Assume that $h$ is continuous and $F$ is non-decreasing. Denote by $Q$ any strictly increasing continuous function from $[c,d]$ onto $[a,b]$, then the following equality holds:
$$\int_c^{d} h(Q(x)) \mathrm{d}F(Q(x)) = \int_a^{b} h(x) \mathrm{d}F(x).$$ 
\end{theorem}
Herein, all integrals are understood as Riemann-Stieltjes integrals.
Note that if $F$ is strictly increasing and if $Q$ denotes the inverse of $F$, then we arrive at 
\begin{equation}
\int_c^{d} h(F^{-1}(x)) \mathrm{d}x = \int_a^{b} h(x) \mathrm{d}F(x).
\end{equation} 
Further, if $F$ denotes a strictly increasing distribution function of a continuous random variable $Y$, then the right hand side denotes the expectation of $h(Y)1_{a<Y<b}$, the inverse $Q$ equals the quantile function and by taking limits one obtains
\begin{theorem}\label{substitution}
Let $F$ denote a strictly increasing continuous distribution function of a random variable $Y$ and let $Q$ denote the corresponding quantile function.
Then for any continuous $h$ such that $\mathbb{E}(|h(Y)|)<\infty$, we have
$$\mathbb{E}(h(Y))=\int_{0}^1 h(Q(q))\mathrm{d}q.$$
\end{theorem} 
In fact, the above result is only a very specialized form of a more general result (see e.g. ch. 2 of \citet{barndorff2015change}).
In particular, it may also be shown that it holds for nonnegative functions $h$ which are not necessarily
continuous. \\

This result can be used to compute moments of the random variable $Y$ by taking $h(x)=x^k$ and also central moments by taking $h(x)=(x-\mu)^k$.
The following proposition contains the general formula for the moments as well as the specific formulas for the two important special cases of linear and logarithmic difficulty functions.
\begin{theorem} \label{meanmoments} Means and central moments

One obtains for strictly increasing difficulty functions
\begin{align}
&\E(Y_{pi})=\int_0^1 \delta_i^{-1}(\theta -F^{-1}(1-q)/\alpha_i ) dq,\\
&\E(Y_{pi}-\mu_{pi})^k = \int_0^1 (\delta_i^{-1}(\theta -F^{-1}(1-q)/\alpha_i) -\E(Y_{pi}))^k dq.
\end{align}

For linear functions, $\delta_i(y)=\delta_{0i}+\delta_i(y)$ with inverse $\delta^{-1}(x)=({x-\delta_{0i}})/{\delta_i}$, one obtains
\begin{align}
&\E(Y_{pi})= \frac{\theta_p-\delta_{0i}}{\delta_i}-\frac{\mu_F}{\alpha_i \delta_i}, \quad
\E(Y_{pi}-\mu_{pi})^k=\int_0^1 \left(\frac{\mu_F-F^{-1}(1-q)}{\alpha_i \delta_{i}}\right)^k dq,
\end{align}
where $\mu_F=\int_0^1 F^{-1}(1-q)dq$ equals the expectation of a random variable with distribution function $F$ according to Proposition \ref{substitution}.

For logarithmic functions one obtains
\begin{align}
&\E(Y_{pi})= c_i\exp\left(\frac{\theta_p-\delta_{0i}}{\delta_i}\right), \quad
\E(Y_{pi}-\mu_{pi})^k= c_{ii}\exp\left(k\frac{\theta_p-\delta_{0i}}{\delta_i}\right),
\end{align}
$c_i=\int_0^1 \exp(-F^{-1}(1-q)/{(\alpha_i \delta_{i}}))dq$, $c_{ii}=\int_0^1 (\exp(-F^{-1}(1-q)/{(\alpha_i \delta_{i}}))-c_i)^kdp$.
\end{theorem}

\begin{proof} 
Recall that the quantile function of a TM model is given by
$$Q(q)=\delta_i^{-1}(\theta -F^{-1}(1-q)/\alpha_i )$$
For a continuous random variable $Y$ with density $F$ and quantile function $Q_Y$ the expectation of a strictly transformation $h(Y)$ can be obtained via Proposition \ref{substitution} by 
\begin{equation}\label{eq:trans}
\E(h(Y))=\int h(y)dF(y)= \int_0^{1} h(Q_Y(q))dq.
\end{equation}
Thus, for strictly increasing difficulty functions and $h(y)=y$ one obtains
\[
\E(Y)=\int_0^{1} \delta_{i}^{-1}(\theta_p - F^{-1}(1-q)/\alpha_i).
\]
With $h(y)=(y-\E(Y_{pi}))^k$ one obtains the corresponding centered moments. 

The inverse of the linear difficulty function $\delta_{i}(y)= \delta_{0i}+ \delta_{i}y $ is given by $\delta_{i}^{-1}(x)=(x-\delta_{0i})/\delta_{i}$ yielding
\[
\E(Y)=\int_0^{1} \frac{\theta_p - F^{-1}(1-q)/\alpha_i-\delta_{0i}}{\delta_{i}}dq=\frac{\theta_p - \delta_{0i}}{\delta_{i}}-\frac{E_F}{\alpha_i\delta_{i}},
\]
where $E_F=\int_0^{1} F^{-1}(1-q)$ is the expectation of a random variable with distribution function $F(.)$. For the central moments one has
\begin{align*}
\E(Y_{pi}-\mu_{pi})^k&=\int_0^{1} \left(\frac{\theta_p - F^{-1}(1-q)/\alpha_i-\delta_{0i}}{\delta_{i}}-\frac{\theta_p - \delta_{0i}}{\delta_{i}}+\frac{E_F}{\alpha_i\delta_{i}}\right)^k dq\\
&= \frac{\int_0^1(\mu_F-F^{-1}(1-q))^k dq}{(\alpha_i \delta_{i})^k}.
\end{align*}
For $k=2$ $\var_F=\int_0^1(\mu_F-F^{-1}(1-q))^k dq$ is the variance of  a random variable with distribution function $F(.)$.

The corresponding functions for logarithmic difficulty functions are obtained by using the inverse function $\delta_{i}^{-1}(x)=\exp((x-\delta_{0i})\delta_i^{-1})$. 
\end{proof}

%
%
%

\begin{theorem}\label{transform_delta}
Given a TM, the transformed variable $\delta(Y_{pi})$ has the same distribution as the random variable 
$$\theta_p-\frac{Y_0}{\alpha_i}$$
with $Y_0 \sim F$
\end{theorem}
\begin{proof}
Let $A$ denote a Borel-measurable set in $\mathbb{R}$. According to the remark following Proposition \ref{substitution}, we can invoke the result of the proposition with nonnegative functions. We choose $h(y):=I_{\delta(y) \in A}$, wherein $I_B$ denotes the indicator function of a set $B$.
Then we get (abbreviating $\theta:=\theta_p$):
\begin{align*}
P(Y_{pi}\in A)=&\mathbb{E}(h(Y_{pi}))=\int_0^1 I_{\delta(Q(q)) \in A} \mathrm{d}q = \int_0^1 I_{\theta-F^{-1}(1-q)\alpha^{-1} \in A} \mathrm{d}q \\
=& \int_0^1 I_{\theta-F^{-1}(q)\alpha^{-1} \in A} \mathrm{d}q = \int I_{\theta - z \alpha^{-1} \in A} \mathrm{d}F(z) = P(\theta-Y_0 \alpha^{-1} \in A).
\end{align*}
\end{proof}

Using the above Proposition, we may immediately deduce the following:
\begin{theorem} \label{try} 
For the thresholds model $P(Y_{pi} > y|\theta_p,\alpha_i,\delta_{i}(.))=F(\alpha_i(\theta_p-\delta_{i}(y)))$ with strictly increasing continuous distribution function $F(.)$ and corresponding density $f(y) = \partial F(y)/\partial y$ one obtains
\begin{align}
&\E(\delta_{i}(Y_{pi})) = \theta_p -  \mu_F/\alpha_i\\
&\var(\delta_{i}(Y_{pi})) = \var_F/ \alpha_i^2,
\end{align}
where $\mu_F=\int y f(y)dy$ is the expectation corresponding to distribution function $F(.)$, and $var_F =\sigma_F^2=\int (y-E_F)^2f(y)d y$ is the variance linked to $F(.)$
If the difficulty function is parameterized by $\delta_{i}(y)= \delta_{0i}+ \delta_i g(y)$, $\delta_i \ge 0$, one obtains for the expectation and the variance
of $g(.)$
\begin{align}
&\E(g(Y_{pi})) = (\theta_p - \delta_{0i} - E_F/\alpha_i)/\delta_i,\\
&\var(g(Y_{pi})) = var_F /(\alpha_i\delta_i)^2.
\end{align}
\end{theorem}

\begin{proof}
According to Proposition \ref{transform_delta}, $\delta_{i}(Y_{pi})$ is distributed as $\theta_p-\frac{Y_0}{\alpha_i}$. Hence, the expecated value  and the variance are given by 
\begin{align*} 
\mathbb{E}(\delta_{i}(Y_{pi}))=&\theta_p- \frac{\mathbb{E}(Y_0)}{\alpha_i}=\theta_p- \frac{\mu_F}{\alpha_i}, \\
\var(\delta_{i}(Y_{pi}))=& \frac{1}{\alpha^2_i}\var(Y_0)=\frac{1}{\alpha^2_i}\var_F.
\end{align*}
The formulas for the special case $\delta_i(y):=\delta_{0i}+\delta_i g(y)$ follow by using the above formulas and by noting that
$$g(Y_{pi})=\frac{\delta_{i}(Y_{pi})-\delta_{0i}}{\delta_i}.$$
\end{proof}
%
%
%
%
 %
%
%


\end{document}